\documentclass[paper]{JHEP}
\usepackage{latexsym,amssymb}
\usepackage{epsfig}
\newcommand{\CMP}[1]{Comm.\ Math.\ Phys.\ {\bf #1}}
\newcommand{\NP}[1]{Nucl.\ Phys.\ {\bf #1}}
\newcommand{\eqn}[1]{(\ref{#1})}
\newcommand{\secn}[1]{Section~\ref{#1}}
\newcommand{\bra}[1]{\langle{#1}|}
\newcommand{\ket}[1]{|{#1}\rangle}
\newcommand{\dket}[1]{|{#1}\rangle\hskip -3pt\rangle}

\newcommand{\dbra}[1]{\langle\hskip -3pt\langle{#1}|}

\newcommand{\eq}[1]{Eq.~(\ref{#1})}

\def\beq{\begin{equation}}
\def\eeq{\end{equation}}
\def\beqa{\begin{eqnarray}}
\def\eeqa{\end{eqnarray}}
\newcommand{\sect}[1]{\setcounter{equation}{0}\section{#1}}

\newcommand{\EQ}{\begin{equation}}
\newcommand{\EN}{\end{equation}}
\newcommand{\bea}{\begin{eqnarray}}
\newcommand{\ena}{\end{eqnarray}}
\newcommand{\eea}{\end{eqnarray}}

\renewcommand{\a}{\alpha}
\renewcommand{\b}{\beta}

\renewcommand{\d}{\delta}
\newcommand{\e}{\epsilon}

\def\one{{\hbox{ 1\kern-.8mm l}}}

\def\ii{{\rm i}}
\def\ee{\mathrm{e}}

\def\acomm#1#2{\left\{ #1, #2\right\}}

\def\sp#1#2{{\scriptstyle{#1\brack#2}}}
\def\tr{{\rm tr\,}}
\def\Tr{{\rm Tr}}

\newlength{\bredde}
\def\slash#1{\settowidth{\bredde}{$#1$}\ifmmode\,\raisebox{.15ex}{/}
\hspace*{-\bredde} #1\else$\,\raisebox{.15ex}{/}\hspace*{-\bredde} #1$\fi}

\newsavebox{\uuunit}
\sbox{\uuunit}
    {\setlength{\unitlength}{0.825em}
     \begin{picture}(0.6,0.7)
        \thinlines
        \put(0,0){\line(1,0){0.5}}
        \put(0.15,0){\line(0,1){0.7}}
        \put(0.35,0){\line(0,1){0.8}}
       \multiput(0.3,0.8)(-0.04,-0.02){12}{\rule{0.5pt}{0.5pt}}
     \end {picture}}

\def\a{\alpha}
\def\b{\beta}

\def\d{\delta}
\def\e{\varepsilon}

\def\g{\gamma}

\def\n{\nu}

\def\r{\rho}                    
\def\s{\sigma}                  
\def\t{\tau}

\title{Orbifold boundary states from Cardy's condition}
\author{Marco Bill\'o\thanks{\email{billo@to.infn.it}}\\ 
Dipartimento di Fisica Teorica, Universit\`a di Torino and \\
I.N.F.N., Sezione di Torino, via P. Giuria 1, I-10125, Torino, Italy}
\author{Ben Craps\thanks{\email{craps@theory.uchicago.edu}}\\
Enrico Fermi Institute, University of Chicago, 5640 S. Ellis Av., 
Chicago, IL 60637, USA\thanks{Present address} \\
Instituut voor Theoretische Fysica, Katholieke Universiteit Leuven, 
B-3001 Leuven, Belgium}
\author{Frederik Roose\thanks{\email{Frederik.Roose@fys.kuleuven.ac.be}}\\
Instituut voor Theoretische Fysica, Katholieke Universiteit Leuven, 
B-3001 Leuven, Belgium\\
Institut Henri Poincar\' e, 11 rue Pierre et Marie Curie, F-75231 Paris 
Cedex 05, France}
\abstract{Boundary states for D-branes at orbifold fixed points are constructed in close
analogy with Cardy's derivation of consistent boundary states in RCFT. 
Comments are made on the interpretation of the various coefficients in the 
explicit expressions, and the relation between fractional branes and wrapped 
branes is investigated for $\mathbb{C}^2/\Gamma$ orbifolds. The boundary 
states are generalised to theories with discrete torsion and a new check is
performed on the relation between discrete torsion phases and projective 
representations.}
\preprint{DFTT 38/2000\\ EFI-2000-44  
\\ KUL-TF-2000/27\\ IHP-2000/16}
\begin{document}
\sect{Introduction}
String theory is often well-behaved on classically singular spaces. As an 
example, consider type IIA string theory compactified on a K3 manifold. When a 
two-sphere of the K3 collapses to zero volume, the resulting space is 
singular. Duality with the heterotic string
relates this  singularity to the appearence of an extended gauge symmetry
\cite{Hull:1995ys,Witten:1995ex}.  The additional massless gauge
bosons are naturally interpreted as D2-branes wrapping the vanishing 
two-sphere. This phenomenon is non-perturbative in string theory,
\emph{i.e.}, it is 
beyond reach of conformal field theory (CFT).
\par
Close to the singularity, the degenerate K3 manifold looks like  
$\mathbb{C}^2/\Gamma$, with $\Gamma$ a discrete subgroup of SU(2). String
theory on $\mathbb{C}^2/\Gamma$ can be described by an orbifold
CFT~\cite{Dixon:1985jw,Dixon:1986jc}. The  difference with the  previous
paragraph, where there was no CFT description, is that the orbifold 
automatically breaks the extended gauge symmetry by a $B$-flux through the 
vanishing cycles \cite{Aspinwall:1995zi,Nahm:1999ps}. 
The wrapped D2-branes appear as fractional  D0-branes in the orbifold
theory~\cite{Douglas:1997xg,Diaconescu:1998br}.
\par
The boundary state formalism
\cite{Polchinski:1988tu,Callan:1987px,Callan:1988wz,Ishibashi:1989tf}  (see,
for instance, \cite{DiVecchia:1999rh,DiVecchia:1999fx,Craps:2000zr,
Gaberdiel} for recent reviews)  is a powerful framework for studying  D-branes. D-branes were
originally introduced in string theory as hyperplanes  on which open strings
can end \cite{Dai:1989ua,Horava:1989vt,Polchinski:1995mt}. Boundary states
provide a closed string  description of D-branes. Consistency with the open
string description  \cite{Pradisi:1989xd} leads to severe constraints on the
possible boundary states in a given string theory. These constraints were
analysed by Cardy \cite{Cardy:1989ir} for a class of rational  conformal field
theories (RCFT), and have become known as ``Cardy's  condition''. A careful
study of this condition allowed Cardy to explicitly  construct a set of
consistent boundary states. 
\par
In this paper we derive boundary states for fractional D-branes at  an orbifold
fixed point in close  analogy with Cardy's construction. The result generalises
the boundary states  of 
\cite{Diaconescu:2000dt,Gaberdiel:2000ch,Takayanagi:2000rd,Roose:2000ay,
Billo:1999nf}.  The
main advantage of our approach is that it makes the origin of the various
coefficients in the expressions for the boundary states manifest.  For
instance, for $\mathbb{C}^2/\Gamma$ orbifolds, typical sine factors arise from 
the modular transformation properties of chiral blocks. We show how these 
transformation properties are consistent with closed string one-loop modular 
invariance for both compact and non-compact orbifolds. 
\par
Some subtleties appear in the Cardy-like construction of the fermionic parts 
of boundary states, even in flat space. Using observations in 
\cite{Lerche:1987cx,Kawai:1987ah,Bianchi:1990yu} we show
that the analogy with Cardy's construction is pretty close, though. For 
instance, the boundary states for ``electric'' and ``magnetic'' type 0 D-branes
\cite{Bergman:1997rf,Klebanov:1999yy,Billo:1999nf} come out very naturally.
\par
We elaborate on the correspondence between fractional and wrapped branes for 
$\mathbb{C}^2/\Gamma$ orbifolds. For more general orbifolds, the wrapped brane 
picture would only be valid when the orbifold is blown up to large volume, but
in the case we are considering there is enough supersymmetry for the picture to
make sense for the undeformed orbifold as well. We determine the explicit basis
transformation that takes one from the twisted Ramond-Ramond potentials to the
potentials associated to the vanishing cycles.
\par
Recently \cite{Sharpe:2000ki,Gaberdiel:2000fe,Aspinwall:2000xs,Aspinwall,Feng} there has been a lot of interest in orbifolds with discrete  torsion~%
\cite{Vafa:1986wx}. Discrete torsion means that the action of the
orbifold group in twisted sectors is modified by certain phase factors,
``discrete torsion  phases''. In \cite{Douglas:1998xa} it was argued that
D-branes in these theories are  associated to projective representations of the
orbifold group, where the  factor system of the projective representations is
related to the discrete torsion phases. We write down boundary states for
fractional branes at a fixed point of an orbifold with discrete torsion
(extending results of  \cite{Diaconescu:2000dt}). We use these boundary states
to check the consistency of the  relation between factor systems and discrete
torsion phases for D-branes  localised at a fixed point.%
\footnote{Note, however, that our analysis is only valid for D-branes
localised at a fixed point. For more general D-branes there is no such 
straightforward relation between
discrete torsion phases and projective representations.\cite{Gaberdiel:2000fe}}
\par
The paper is organised as follows. Section~\ref{sec:cardy} contains a brief
review of  Cardy's condition and the construction of consistent boundary states
in RCFT. In Section~\ref{sec:dbraneflat} we construct boundary states for
D-branes in flat space.  Section~\ref{sec:Dbofp} deals with D-branes at
orbifold fixed points. First we briefly review the general structure of
orbifold theories. Next we study modular  transformation properties of twisted
Virasoro characters and write down  modular invariant partition functions for
both compact and non-compact  orbifolds. Then we construct boundary states for
fractional branes at orbifold fixed points. For $\mathbb{C} ^2/\Gamma$
orbifolds we comment on the relation to  McKay correspondence \cite{mckay} and
on the relation with wrapped branes.  In section~\ref{sec:discretetorsion} we
study D-branes at fixed points of discrete torsion orbifolds.  
Appendix~\ref{bosbsflat} discusses the simple example of the free
bosons in flat space. 
Appendix~\ref{app:chirblocks} contains details on chiral blocks and their
modular transformation properties. In Appendix~\ref{app:useful} we have
collected some  information about theta functions and discrete groups.
\sect{Cardy's condition}
\label{sec:cardy}
In this section, Cardy's construction of consistent boundary states is briefly
reviewed. The original prescription was derived in the context of a rational
conformal field theory on  a cylinder. It may be considered as a direct
implementation of open-closed consistency. We closely follow the original
derivation in \cite{Cardy:1989ir}. As it turns out, the procedure can be generalised
in specific instances of non-rational CFTs. We defer the details of that to the
following sections, however.
\subsection{Rational CFT}
Rational CFT is a realisation of holomorphic and anti-holomorphic symmetry 
algebras ${\cal A}_L$  and ${\cal A}_R$, both containing the Virasoro algebra. 
The number of primaries in a rational CFT is finite. We label these primaries
by an index $i$. The Virasoro characters in the corresponding irreducible
modules are 
\beq\label{holchar}
\chi_i(q)=\Tr_iq^{L_0-\frac c{24}}~,
\eeq
where $q=e^{2\pi \ii \tau}$, with $\tau$ the modular parameter. We shall only
consider purely  imaginary $\tau$ in this paper. 
\par
One identifies a sub-algebra ${\cal A}$ of both ${\cal A}_L$ and ${\cal A}_R$, 
and denotes the corresponding generators by $W^{(r)}$ and $\tilde W^{(r)}$. 
The choice of ${\cal A}$ will largely determine what kind of D-branes (open
strings, boundary states) one wishes to keep in the theory. At world-sheet
boundaries the generators of the holomorphic and the anti-holomorphic
embeddings of ${\cal A}$ are related  through gluing  conditions. In the rest
of this section, it is assumed that the ``preserved'' symmetry algebra ${\cal
A}$ coincides with the  algebras ${\cal A}_L$ and ${\cal A}_R$, which are thus
isomorphic.  In this section we only consider theories with diagonal partition
function on the torus, \emph{i.e.},
\beq
Z(q,\bar q)=\sum_j \chi_j(q)\left(\chi_j(q)\right)^*~.
\eeq
\par 
Consider the cylinder amplitude with boundary conditions labeled by
$\alpha$ and $\beta$. In the loop channel, one has
\begin{equation}\label{amploop}
Z_{\alpha\beta}= \sum_i n_{\alpha\beta}^i \chi_i(q)\ ,
\end{equation}
where the integers $n_{\alpha\beta}^i$ denote the multiplicities of the
representations $i$ running in the loop. In the tree channel, this amplitude
reads
\begin{equation}\label{amptree}
Z_{\alpha\beta}= \bra \alpha \tilde q^{\frac 1 2 (L_0 + \tilde L_0 -\frac c {12})} \ket \beta\ ,
\end{equation}
where $\tilde q = e^{-2\pi \ii / \tau}$.
In the latter equation, the boundary states $\bra \a$ and $\ket\b$
impose the boundary condition 
\begin{equation}
\left(W^{(r)}_{n} - (-)^s \Omega(\tilde W^{(r)}_{-n}) \right) = 0 
\end{equation}
on the Fourier modes of the symmetry generators; that is, 
\begin{equation}\label{BSbc}
\left(W^{(r)}_{n} - (-)^s \Omega(\tilde W^{(r)}_{-n}) \right) \ket \beta = 0 ~,
\end{equation}
where $s$ is the spin of $W^{(r)}$ and $\Omega$ an automorphism\footnote{%
For instance, in the case of a free boson, choosing $\Omega$ the trivial
automorphism corresponds to Neumann boundary conditions, while a non-trivial one 
gives Dirichlet conditions (see Appendix \ref{bosbsflat}).} of the preserved
algebra $\mathcal{A}$. In the remaining of this section we assume for simplicity
that $\Omega$ is the trivial automorphism.
\par
The way to proceed from here is to choose a convenient basis of solutions to
\eq{BSbc}, the so-called Ishibashi states~\cite{Ishibashi:1989tf}. The
consistent boundary states are then built as particular linear combinations of
the Ishibashi states. Consider a highest weight representation $j$ of ${\cal
A}_L$ and the corresponding (isomorphic) representation $\tilde j$ of ${\cal
A}_R$. The states of $j$ are linear combinations of states of the form
$\prod_IW^{(r_i)}_{-n_I}\ket{j;0}$, where the  $W^{(r_i)}_{-n_I}$ are lowering
operators and $\ket{j;0}$ is the highest weight state. Denote the elements of
an orthonormal basis of the representation $j$ by $\ket{j;N}$ and the
corresponding basis of $\tilde j$ by $\widetilde{\ket{j;N}}$. In terms of the
anti-unitary operator $U$ defined by
\begin{equation}
U \widetilde{\ket {j;0}} = \widetilde{\ket{j;0}}^*\ ;\ U\tilde W^{(r_I)}_{-n_I} U^{-1} =
(-)^{s_{r_I}} \tilde W^{(r_I)}_{-n_I} \ ,
\end{equation}
the states
\begin{equation}\label{ishibashi}
\dket j \equiv \sum_N \ket {j;N} \otimes U \widetilde{\ket{j;N}}
\end{equation}
solve \eq{BSbc} (for $\Omega=1$). These are the Ishibashi states. 
\subsection{Cardy's condition}
The consistent boundary states are linear combinations of the Ishibashi
states  \eq{ishibashi}:
\beq\label{combinationishi}
\ket\a=\sum_j B_{\a}^j \dket j~.
\eeq
We can now rewrite \eq{amptree} as
\begin{equation}\label{amptreeishi}
Z_{\alpha\beta} = \sum_j (B^j_\a)^* B^j_\b\chi_j(\tilde q)\ ,
\end{equation}
where
\beq\label{ishichar}
\chi_j(\tilde q)=
\dbra j \tilde q^{\frac 1 2 (L_0 + \tilde L_0 -\frac c {12})}\dket j=
\Tr_j\tilde q^{L_0-\frac c{24}}~,
\eeq
which is consistent with \eq{holchar}. In deriving \eq{amptreeishi} we have
used the fact that the Ishibashi states are orthogonal in the sense that
\beq\label{ortho}
\dbra {j'} \tilde q^{\frac 1 2 (L_0 + \tilde L_0 -\frac c {12})}\dket j=0
\ {\rm if}\ j\neq j'~.
\eeq
\eq{amploop} is transformed to the tree channel by the modular $S$ 
transformation:
\begin{equation}\label{Ztransf}
Z_{\alpha\beta} = \sum_{i,j} n_{\alpha\beta}^i S_i^j \chi_j(\tilde q)\ .
\end{equation}
Demanding equality of \eq{amptreeishi} and \eq{Ztransf} eventually yields
\begin{equation}\label{cardycond}
\sum_i S_i^j n_{\alpha\beta}^i = (B^j_\a)^* B^j_\b\ ,
\end{equation}
at least if no two representations of the holomorphic algebra have the same
Virasoro character. \eq{cardycond} is called Cardy's equation. The requirement
that the multiplicities $n_{\alpha\beta}^i$ be non-negative integer numbers is
a strong condition on the coefficients $B^j_\a$. This constraint is non-linear:
multiplying a consistent boundary state by a non-integer number will
generically not yield a consistent boundary state.
\subsection{Cardy's solution}
\label{sec:cardysol}
In \cite{Cardy:1989ir}, Cardy gave a  solution to his equation
\eq{cardycond}.  The first consistent boundary that was constructed, is
$\ket {\bf 0}$. It is determined by the requirement that $n_{00}^i =
\delta_0^i$, that is, the only representation running in the loop channel is
the identity representation. From \eq{cardycond}, such a state satisfies
\begin{equation}\label{zero} |B_0^j|^2 = S_0^j\ .
\end{equation}
The entries $S_0^j$ of the modular transformation
matrix are positive \cite{Cardy:1989ir}, so \eq{zero} is consistent. 
\eq{zero} implies
\begin{equation}
\label{zeroS}
\ket {\bf 0} = \sum_j \sqrt{ S_0^j} \dket j 
\end{equation}
(up to the relative phases of the coefficients, which are not fixed by these
considerations).
Similarly, one defines boundary states $\ket {\bf l}$, with the property that 
$n_{0l}^i = \delta_l^i$,
for every primary $l$. Using \eq{cardycond}, \eq{zero} and the fact that
$S_0^j>0$, these states  are seen to be
\begin{equation}\label{coeffS}
\ket{\bf l} = \sum_j \frac {S_l^j}{\sqrt{ S_0^j}} \dket j \ .
\end{equation}
In Cardy's solution, the multiplicities $n_{\alpha\beta}^i$ coincide with the
fusion rule coefficients of the algebra ${\cal A}$. The boundary states
\eq{coeffS} solve \eq{cardycond} because of Verlinde's formula \cite{verlinde,
Cardy:1989ir}.
\sect{D-branes in flat space}
\label{sec:dbraneflat}
Cardy's construction of consistent boundary states can be made explicit is the 
important example of the CFT of free bosons. Since this is well-known, we 
defer most of the discussion to Appendix~\ref{bosbsflat}. Implementing
Cardy's construction for free fermionic theories is somewhat less trivial.
We shall show that, despite some subtleties, it is possible to follow Cardy's
prescription rather closely. This proves to be an
elegant way to construct directly the type II or type 0 boundary states
satisfying the requirements of open-closed consistency. 
We also briefly discuss
the way in which the proper normalisations of the boundary states in string
theory (as opposed to CFT) are determined.   
\subsection{Boson boundary state}
The basics of the discussion of boundary states in free
bosonic theories in terms of Cardy's condition were given already
in~\cite{Recknagel:1998sb,Fuchs:1998fu}. In this subsection we give a very
brief summary. For completeness, and also in 
order to establish notations and conventions,
details are collected in Appendix~\ref{bosbsflat}. 

On each spacetime coordinate we can impose either Neumann or Dirichlet
boundary conditions, 
corresponding to a trivial ($\Omega=1$) or non-trivial ($\Omega=-1$) choice of
the automorphism $\Omega$.
Correspondingly, the boundary states can be
of Neumann or Dirichlet type in each spacetime direction. 
As an example, the consistent boundary state describing a D-brane localised
at a position $x$ along a circle with radius $R$ is given by
\begin{equation}
\ket{x}_{D} = \left(\frac{\sqrt{\alpha '}}{\sqrt 2 R }\right)^{1/2}
\sum_{k \in \mathbb{Z}} \ee^{\ii x \frac k R}\dket{(k,0)}_D\ ,
\end{equation} 
where the Ishibashi states $\dket{(k,0)}_D$ take the form
\begin{equation}
\dket{(k,0)}_D =
\exp\left(\sum_{n = 1}^{\infty}\frac 1 n \alpha_{-n}\tilde\alpha_{-n}
\right)\ \ket{(k,0)}~.
\end{equation}

\paragraph{String theory}
The introduction of two boundaries at fixed positions necessarily introduces a
dimensionful  scale in the theory. In the resulting CFT on a strip, this scale
is the width of the strip. The context where Cardy first derived his condition
and solutions was in CFT applications to statistical mechanical systems, where
the scale has a physical significance. As such,  equality is demanded of
expressions \eq{amptreeishi} and \eq{Ztransf} on a {\em fixed} cylinder. This
situation is to be contrasted with string theory. Let us adopt the
path-integral point of view for a while. Given a topology, the prescription to
obtain the string amplitude  is then that one should integrate over all metrics
and divide out the volume of the gauge  group, \emph{i.e.}, the (Diff x Weyl)
group. In the present case, this implies the presence of an integral over the
modulus of the cylinder in one-loop open  string amplitudes. 
In the closed string channel, this is reflected by the integral in the closed 
string propagator below. We shall now indicate how equality of the amplitudes 
in the open and closed string channels fixes the normalisation of the boundary 
states.
\par
Let us consider strings propagating in flat space. For boundary conditions
$1,2$  corresponding to parallel $p$-branes located at $y_1$, $y_2$, we have 
\beq
\label{1stringbs1}
\mathcal{Z}_{1,2} =  V_{p+1} 
\int_0^{\ii\infty} {d\tau\over \tau} \, Z^{(d)}_{1,2}(q)~,
\eeq
where $q=\exp(2\pi\ii\tau)$, $d=26$ for the bosonic string or $d=10$ for the 
superstring and $Z^{(d)}_{1,2}(q)$ is the standard open string partition
function   in $d$ flat directions, of which $p+1$ are Neumann  (see for
instance \cite{polchinski}), with the prescribed boundary conditions. Notice
that we included in $Z^{(d)}_{1,2}(q)$ the integration over the momentum along
the  Neumann directions. The partition function $Z^{(d)}_{1,2}(q)$ can be
expressed in terms of the Neumann and Dirichlet open string characters
introduced in the previous section, in the decompactified limit $R\to\infty$;
notice however that ghost contributions effectively cancel those of two
Dirichlet directions.  In the superstring case, the contributions of the
fermionic sectors, to be considered shortly in \secn{sec:fermbsflat}, and of
super-ghosts, must obviously be added.  Upon modular transformation, the open
string amplitude $\mathcal{Z}_{1,2}$  has to be reinterpreted in terms of
closed string propagation between boundary states:
\beq
\label{1stringbs2}
\bra{B_1} \frac{\a'}{4\pi}\int_{|z|<1} \frac{d^2 z}{|z|^2} z^{L_0-a} \bar
z^{\tilde L_0 -\tilde a}\, \ket{B_2}~.
\eeq     
Here we used the canonical normalization of the closed string propagator.  The
boundary states $\ket B$ that do the job are given by the product of 
consistent 
boundary states of Neumann type, \eq{thetaNbs}, for $p+1$ directions, 
and of states of Dirichlet type, \eq{xDbs}, for $d-(p+1)-2$ directions (the 
$-2$ accounts for the ghost contributions), times an overall normalisation. 
In the decompactified  limit for all directions, the appropriate overall 
normalisation\footnote{%
For commodity, we collect in this normalisation also the pre-factors of powers of
$\alpha'$ and $2$ that in the previous section appeared in the 
consistent states \eq{thetaNbs}, \eq{xDbs}.}
is given (see for instance \cite{DiVecchia:1999rh}) by
\beq
\label{1stringbs3}
N_p = \frac{\sqrt{\pi}}{2}\, 2^{\frac{10 - d}{4}}\,
(2\pi\sqrt{\a'})^{\frac{d-4}{2} -p}~.
\eeq
The corresponding tension of the D$p$-brane, as computed from the
graviton-dilaton exchange in the field-theory limit of the closed string
amplitude \cite{Polchinski:1995mt} is $T_p/\kappa$, with $T_p = 2 N_p$
and $\kappa$ being the gravitational coupling constant. 
\subsection{Fermion boundary state}
\label{sec:fermbsflat}
Let us next focus on the world-sheet fermions. In principle, these may be
treated in the covariant formulation \cite{Billo:1998vr}, similarly to the
bosons above.  For simplicity, however, we opt for a light-cone treatment of
the fermionic sector. Apart from one subtle point explained shortly, there is
no obvious gain for our purposes in the covariant formulation. The fermions
thus realise the SO$(8)$ affine algebra ${\cal A}$ at level 1. In both the NS-
and R-sectors, the unprojected Fock space decomposes into two irreducible
modules  ${\cal H}^\pm$ of ${\cal A}$, according to the GSO projection.  The
corresponding four inequivalent representations of the $\mathrm{SO}(8)$ algebra
${\cal A}$ are the singlet $(o)$ and the vector $(v)$ in the NS sector, the
spinor $(s)$ and conjugate spinor $(c)$ in the R sector. Their characters are
$\chi_a(q) = {\rm Tr}_a(q^{L_0-{1\over 6}})$, where  $a=o,v,s,c$.
\par
We put an extra minus sign in the definition of the spinorial characters
$s,c$ compared to their natural definition as SO$(8)$ characters. This
peculiar choice actually implements the space-time spin-statistics
\cite{Kawai:1987ah,Lerche:1987cx,Bianchi:1990yu}. With this choice, the
SO$(2n)_1$ characters are given explicitly by
\bea
\label{ven1}
\chi_v  = {1\over 2} \left[\left({\theta_3\over \eta}\right)^n - 
\left({\theta_4\over \eta}\right)^n\right]~,\hskip 0.2cm & &
\chi_o  = {1\over 2} \left[\left({\theta_3\over \eta}\right)^n + 
\left({\theta_4\over \eta}\right)^n\right]~,
\nonumber\\
\chi_s  = -{1\over 2} \left[\left({\theta_2\over \eta}\right)^n + 
\left({\theta_1\over \eta}\right)^n\right]~,\hskip 0.2cm & &
\chi_c  = -{1\over 2} \left[\left({\theta_2\over \eta}\right)^n - 
\left({\theta_1\over \eta}\right)^n\right]~,
\eea
where, in fact, $\theta_1(\tau)$ is zero. For now, our primary interest is
in the $n=4$ case.
\par 
The modular $T$ and $S$ transformations are encoded in the following matrices
acting on the character vector $\chi_a$ (ordered as $v,o,s,c$) 
\bea
\label{ven2}
T_{(2n)} & = & \mathrm{diag}\,(-\ee^{-{n\pi\ii\over 12}},\ee^{-{n\pi\ii\over 12}},
\ee^{{n\pi\ii\over 6}},\ee^{{n\pi\ii\over 6}})~,
\nonumber\\
S_{(2n)} & = & {1\over 2} \left(\matrix{ 1 & 1 & 1 & 1 \cr
                                  1 & 1 & -1 & -1\cr
				  1 & -1 & (-\ii)^n  & -(-\ii)^n\cr
				  1 & -1 & -(-\ii)^n & (-\ii)^n }\right)~.
\eea				    
The form of $S_{(2n)}$ suggests that it is the $v$ class rather than the
singlet class $o$, that plays the role of the identity in the  fusion rules.
This exchange has been explained in the covariant formulation, where one
considers the combined system of SO$(1,9)$ fermions and  super-ghosts
\cite{Lerche:1987cx}. Indeed, thinking of fusion rules, it is logical that a
representation with  even world-sheet spinor number should play the role of the
identity representation.   
\par
The bulk conformal field theory is specified by a choice of modular invariant.
To make the analogy with \secn{sec:cardy} as close as possible, we shall first
consider a diagonal partition function. The torus partition function for this
theory is 
\beq
\label{Ztype0}
Z^{\mathrm{0B}}(q) \propto 
(|\chi_o(q)|^2 + |\chi_v(q)|^2+|\chi_s(q)|^2+|\chi_c(q)|^2)~,
\eeq
\emph{i.e.}, the following (left,right) sectors are kept: 
$(NS+,NS+)\oplus(NS-,NS-)\oplus(R+,R+)\oplus(R-,R-)$. The contribution of the
bosonic string fields completes this expression to the full type 0B torus
partition function.
\par
Starting from this bulk CFT, boundaries may be introduced in string world-sheets. Below
attention will be focused on D9-branes%
\footnote{Of course, strictly speaking it is inconsistent to consider a 
D9-brane without the corresponding anti-brane, because this would lead to a R-R
tadpole. Moreover, for D9 branes (or D8's in type IIA) one actually needs a 
covariant formulation, as the exact cancellation of the ghost-superghost
contributions  against two coordinates takes place for Dirichlet directions.  
This subtlety shows up, for instance, in systems with eight ND directions, 
like the  D0-D8 one~\cite{Billo:1998vr}.}, preserving the full Lorentz, whence
SO$(8)$ invariance in light-cone gauge. The D$p$-branes with $p<9$ are 
obtained by T-duality.
\par
Let us first  construct the Ishibashi states. In both the NS and R sectors, two
gluing conditions are possible:
\begin{equation}
\psi_r = \ii\eta \tilde \psi_{-r}\ ,
\end{equation}
where $\eta = \pm 1$. 
States solving these conditions as in \eq{ishibashi} are 
\begin{equation}
\label{naivestates}
\dket{\sigma;\eta} = \prod_\mu\exp\left[\ii\eta \sum_{r > 0} 
\psi^\mu_{-r} \tilde \psi^\mu_{-r}
\right] \dket{\sigma,\eta;0}\ ,
\end{equation}
with $\sigma = {\rm NS,R}$ indicating the NS-NS or R-R sector and $\mu$ running
on the transverse directions. In the R-R (NS-NS) sector, the mode numbers $r$
are (half-)integer. Also, there is a non-trivial zero-mode part in the R
sector:
\begin{equation}
\label{Rzeromodes}
\dket{{\rm R},\eta;0} = 
{\cal M}_{AB}^{(\eta)}\,\ket{A} \ket{\tilde B}~~, 
\eeq
where
\begin{equation}
\label{bs14}
{\cal M}^{(\eta)} = C\Gamma^0\Gamma^{l_1}\ldots
\Gamma^{l_p} \,\left(
\frac{1+\ii\eta\Gamma_{11}}{1+\ii\eta}\right)~~,
\end{equation}
with $C$ being the charge conjugation matrix and $l_i$ labeling the space 
directions
of the D-brane world volume. The vacuum states $\ket{A}\ket{\tilde B}$ for the
fermionic zero-modes $\psi^\mu_0$ and  ${\tilde \psi}^\mu_0$ transform in the
32-dimensional Majorana representation.  
\par
From the states in \eq{naivestates} a proper set of Ishibashi states is
provided by taking appropriate linear combinations: 
\begin{eqnarray}
\label{ven4}
\dket{v}&=& \frac 1 {2}\left( \dket{{\rm NS},+} - 
\dket{{\rm NS},-} \right) \ ; \label{NS+}\\
\dket{o} &=& \frac 1 { 2}\left( \dket{{\rm NS},+} +
\dket{{\rm NS},-} \right) \ ; \label{NS-}\\
\dket{s} &=& \frac 1 { 2}\left( \dket{{\rm R},+} +
\dket{{\rm R},-} \right) \ ; \label{R+}\\
\dket{c} &=& \frac 1 { 2}\left( \dket{{\rm R},+} -
\dket{{\rm R},-} \right) \label{R-}\ . 
\end{eqnarray}
They are mutually orthogonal states satisfying the type 0B GSO projection.  
The above labelling follows the general convention of \eq{ishichar}; the
corresponding chiral blocks may indeed be verified to be:
\begin{eqnarray}
\label{ishicharferm}
\dbra{m} {\tilde q}^{\frac 1 2 ({L_0 +
\tilde L_0 -\frac c {12}})} \dket{n} = \delta_{mn}\,\chi_m(\tilde q)~,
\end{eqnarray}
with $m,n=o,v,s,c$. 
From the Ishibashi states we wish to construct a set of consistent boundary
states $\ket{a}$, with $a=v,o,s,c$, that is, we are after a consistent set of
coefficients ${B_a^j}$ of \eq{combinationishi}.
These are obtained by Cardy's formula \eq{coeffS}, with the S matrix of
\eq{ven2}. As alluded to before, the fact that $S_v^m = 1/2>0$  for all $m$ 
reflects the
fact that it is the $v$ representation that is the analogue of the identity
representation in RCFT. As such, $\ket v$ is the analogue of the state 
$\ket{\bf 0}$ in \secn{sec:cardysol}. We obtain
\bea
\label{ven5}
\ket{v} & = &  {1\over \sqrt{2}}\sum_m \dket{m}  = 
{1\over\sqrt{2}}(\dket{v} + \dket{o} + \dket{s} + \dket{c})
\nonumber\\
& = & {1\over\sqrt{2}} (\dket{NS,+} + \dket{R,+})~,
\nonumber\\
\ket{o} & = &  \sqrt{2}\sum_m (S_{(8)})_v^m\,\dket{m}  = 
{1\over\sqrt{2}}(\dket{v} + \dket{o} - \dket{s} - \dket{c})
\nonumber\\
& = & {1\over\sqrt{2}} (\dket{NS,+} - \dket{R,+})~,
\nonumber\\
\ket{s} & = &  \sqrt{2}\sum_m (S_{(8)})_s^m\dket{m}  = 
{1\over\sqrt{2}}(\dket{v} - \dket{o} + \dket{s} - \dket{c})
\nonumber\\
& = & {1\over\sqrt{2}} (-\dket{NS,-} + \dket{R,-})~,
\nonumber\\
\ket{c} & = &  \sqrt{2}\sum_m (S_{(8)})_s^m\dket{m}  = 
{1\over\sqrt{2}}(\dket{v} - \dket{o} - \dket{s} + \dket{c})
\nonumber\\
& = & {1\over\sqrt{2}} (-\dket{NS,-} - \dket{R,-})~.
\eea
These states are the type 0B boundary states that may be found in the 
literature \cite{Bergman:1997rf,Klebanov:1999yy,Billo:1999nf}.  The states
$\ket{v}$ and $\ket{o}$ are usually referred to as  electric D9-brane and
anti-D9-brane, respectively, while  $\ket{s}$ and $\ket{c}$ are called magnetic
D9-brane and anti-D9-brane.
\par
From the Cardy construction the integer multiplicities in the loop channel
$n^c_{ab}$ follow from the Verlinde formula
\cite{verlinde}:
\beq
\label{ven7}
n^c_{ab} = \sum_m {(S_{(8)})_a^m\, (S_{(8)})_b^m\, ((S_{(8)})_c^m)^*
\over (S_{(8)})_v^m} =
2\, \sum_m (S_{(8)})_a^m\, (S_{(8)})_b^m\, (S_{(8)})_c^m~,
\eeq
where the second equality makes use of the explicit form of the matrix
$S_{(8)}$ of  \eq{ven2}.  They correspond to the fusion rules 
\beq
\label{ven8}
v\otimes \a = \a~,~~~ \a\otimes\a = v~,~~~ o\otimes s = c~,~~~ 
o\otimes c = s~,~~~ s\otimes c = o~,
\eeq
that is we have $n_{v\a}^i = \delta^i_\a$, $n_{\a\a}^i=\delta^i_v$,
$n_{oc}^j=\delta^i_s$, and so on. These fusion rules correspond to the algebra
of the conjugacy classes $o,v,s,c$ of SO$(8)$ representations where $o$ and $v$
have been exchanged \cite{Kawai:1987ah,Lerche:1987cx,Bianchi:1990yu}.
\par
Next we want to consider the supersymmetric type IIB theory,
which has the one-loop partition function 
\beq
\label{ZtypeIIB}
Z^{\mathrm{IIB}}(q) \propto |\chi_v(q)+\chi_s(q)|^2~.
\eeq
As is well known (see, \emph{e.g.}, \cite{Billo:1998vr}), of the Ishibashi
states  (\ref{NS-}-\ref{R-}) only $\dket{v}$ and $\dket{s}$ are kept after the
type IIB GSO projection $(1+(-)^{F_L})/2$ $\times (1+(-)^{F_R})/2$. It is no
longer possible to apply Cardy's procedure straightaway; however, the boundary
states ensuring open-closed consistency may be obtained in a similar spirit. 
First reorganise the characters running in the loop channel into the vector 
$\hat\chi_A$ ($A=0,1,2,3$) where
\beq
\label{mar2}
\hat\chi_\a\equiv(\chi_v +\chi_s, \chi_o+\chi_c,\chi_v-\chi_s,\chi_o-\chi_c)
\eeq
and reorder those in the tree channel into  $\hat\chi_M =
(\chi_v,\chi_s,\chi_o,\chi_c)$. With respect to these new bases, the  $S$
modular transformation is $\hat\chi_A(q) =  \hat S_A^M\,\hat\chi_M(\tilde q)$,
with 
\begin{equation}
\label{modS}
\hat S = \left(
\begin{array}{crrr}
1&1&0&0 \\
1&-1&0&0 \\
0&0&1&1 \\
0&0&1&-1
\end{array} \right)\ .
\end{equation}
The matrix $\hat S$ is thus seen to commute with the GSO projection,  which is
just the projector on the upper left block. As such, it makes sense to just
restrict ourselves to the GSO projected Ishibashi states $\dket{\hat M}$, with
$\hat M = v,s$, and introduce consistent states  $\ket{\hat A},(\hat\a=0,1)$
given by a projected version of Cardy's formula \eq{coeffS} (notice that
$\ket{0}$ corresponds to the identity,  with $\hat S_0^{\hat M} = 1>0$):
\bea
\label{mar1}
\ket{0} & = &  \sum_{\hat M} \dket{\hat M} = \dket{v} + \dket{s}
\nonumber\\
& = & {1\over 2}(\dket{NS,+} -\dket{NS,-} + \dket{R,+}+\dket{R,-})~,
\nonumber\\
\ket{1} & = &  \sum_{\hat M} \hat S_1^{\hat M}\dket{\hat M} = 
\dket{v} - \dket{s}
\nonumber\\
& = & {1\over 2}(\dket{NS,+} -\dket{NS,-} - \dket{R,+}-\dket{R,-})~.
\eea
It may be observed that $\ket{0}$ and $\ket{1}$ coincide with the familiar
expressions for IIB D9 and anti-D9 brane boundary states. Amplitudes $Z_{\hat
A\hat B}$ with such boundary conditions are given by
\beq
\label{mar3}
Z_{00} = Z_{11} = \hat\chi_0 = \chi_v +\chi_s~,
\hskip 0.3cm
Z_{01} = Z_{10} = \hat\chi_1 = \chi_o +\chi_c~,
\eeq
yielding the $D9$-$D9$ and $D9$-$\bar{D9}$ amplitudes indeed.
\par 
This Cardy-like derivation of consistent type II theory boundary states
may appear to be somewhat heuristic\footnote{%
For one thing, in these cases the characters are not linearly  independent: 
the one-loop vacuum amplitude for BPS D-branes vanishes due to the familiar
cancellation between the NS and the R sector contributions. Nevertheless,  the
derivation of Cardy-like conditions may be justified by requiring that not only
the  vacuum diagram but also diagrams with vertex operators inserted agree 
between the open and closed string channels.  This is analogous to the use of
unspecialised characters in \cite{Behrend:2000bn}.}. The states that we find in
this way automatically enjoy  a number of nice (physical) properties of
D-branes, though. In particular,   pure RR boundary states are absent, and all
D-branes couple to gravity (see also \cite{Harvey:2000gq}). The nice aspect of the construction is in
displaying these features very explicitly.
\par 
Above, only collections of (parallel) D-branes of the same dimension were
considered. Allowing more general configurations involving D-branes of
different dimensions, mutual consistency will generically impose additional
constraints. As an example, one learns that in type IIA theory with BPS D$p$
branes for even $p$, only non-BPS D$q$-branes with odd $q$ can be added
consistently (see, for instance, \cite{Gaberdiel}). 
\par
Let us finally recall that the complete superstring boundary states  are
obtained by multiplying the bosonic (Neumann or Dirichlet as appropriated to
the various directions) and fermionic boundary states we have been discussing,
with the overall normalization $N_p$ described in  \eq{1stringbs3}, for $d=10$.
The type II boundary states normalised in this way describe  D$p$-branes with
the tension $T_p/\kappa=2N_p/\kappa$  as in \eq{1stringbs3} and a charge
density \cite{Polchinski:1995mt}
\beq\label{RRcharge}
\mu_p = 2\sqrt{2} N_p  = \sqrt{2\pi} (2\pi\sqrt{\a'})^{3-p}
\eeq
with respect to the canonically normalised RR $(p+1)$-form $A_{p+1}$.     
\sect{D-branes at an orbifold fixed point}
\label{sec:Dbofp}
After the flat space D-branes, let us next examine branes in orbifold spaces of
$\mathbb{C}^n/G$ type. Some attention will be paid in particular to the
$\mathbb{C}^{2}/\Gamma$ cases, with $\Gamma$ a discrete subgroup of SU$(2)$.  
The latter orbifolds correspond to blown-down ALE spaces, and are local  models
of degenerate K3 surface geometries. The CFT  description of such orbifolds is
under control, and one primary aim will be to show how a suitably modified
Cardy prescription leads to the correct expressions for consistent D-brane
boundary states.
We will only discuss fractional D-branes localised at the fixed point of the
orbifold.~%
\cite{Polchinski:1997ry,Douglas:1997xg,Douglas:1997de,Diaconescu:1998br}
\par  
In Subsection~\ref{sec:generic} we briefly review a number of general features
of  orbifold conformal field theories. After a more detailed discussion of the
chiral blocks  for orbifolds of free theories and their modular transformation
properties (Subsection~\ref{sec:chiral}),  Subsection~\ref{sec:fractional} 
treats the construction of consistent boundary states for D-branes stuck at
orbifold fixed points. Some comments regarding  the geometrical significance of
these states are made at the end of that subsection.  Finally, we briefly
comment on the relation with  the boundary states constructed in
\cite{Lerche:2000uy,Lerche:2000jb}, where the relation of ALE  spaces with
non-compact Gepner models was exploited.
\subsection{General features of orbifold theories}
\label{sec:generic}
Orbifold theories \cite{Dixon:1985jw} are obtained from a parent conformal
field theory  by modding out a discrete symmetry group $G$. That is, $G$ is a
finite subgroup of the group of endomorphisms of the algebra
$\mathcal{A}_L\otimes\mathcal{A}_R$, preserving the left-right  decomposition
and commuting with the Virasoro algebra \cite{DVVV}. 
\par
The partition function of an orbifold CFT on a torus has the following structure:
\beq\label{partfion}
Z=\frac 1{|G|}\sum_{
\begin{array}{c}{g,h\in G}\\{[g,h]=e}\end{array}
}Z(g,h)\ .
\eeq
More generally, sector-dependent phases $\e(g|h)$, may be introduced in
\eq{partfion}  \cite{Vafa:1986wx}. These numbers adding the $Z(g,h)$ with
relative weights reflect the possibility to turn on discrete torsion. The
discussion of those cases is deferred until \secn{sec:discretetorsion},
however. 
\par
In \eq{partfion}, $Z(g,h)$ represents the partition function evaluated with
boundary conditions twisted by $h$ and $g$ along the $a$- and $b$-cycles of the
torus,  respectively. As $aba^{-1}b^{-1}$ is in the trivial $\pi_1$ homotopy
class, these boundary conditions are only consistent if $g$ and $h$ commute (we
use the notation $[g,h]=ghg^{-1}h^{-1}$), whence the restriction on the sum in
\eq{partfion}.
\par 
From the Hamiltonian 
point of view, the  structure of the orbifold Hilbert space  $\tilde{\cal H}$
is as follows. First, the original untwisted Hilbert space ${\cal H}_0$ is
projected onto its $G$-invariant  subspace. Next, for each $h\in G$, a twisted
Hilbert space ${\cal H}_h$ is introduced,  in which the fundamental fields obey
periodicity conditions twisted by $h$.  As in the untwisted Hilbert space, only
states invariant under the action of the orbifold group $G$ are kept. For
non-abelian orbifold groups, and/or in the presence of different fixed points,
there are additional complications, to which we now turn.
\par
Take a field $\phi(\s,\tau)$ in the sector twisted by $h\in G$ ($\s$ and
$\t$ are the world-sheet space and time coorditates). It satisfies
\beq\label{periodcondg}
\phi(\s+2\pi,\tau)=h\phi(\s,\tau)~,
\eeq  
where $h\phi$ denotes the image of the field $\phi$ under the action of $h$.
Now $G$-invariant states are obtained in two steps. First, any
state in  ${\cal H}_h$ is projected inside ${\cal H}_h$ onto its 
$N_h$-invariant  part, where \beq\label{stabilizer}
N_h=\{g\in G|[g,h]=e\}
\eeq 
is the stabilizer group of $h$. Next one averages by adding the images of such
state in the sectors twisted by the remaining elements in the conjugacy class
of $h$. The final result is a $G$-invariant state in natural correspondence
with the conjugacy class.
\par
How is this related to \eq{partfion}?  The
term $Z(g,h)$ corresponds to
\beq
\label{cop4}
Z(g,h)=\Tr_{{\cal H}_h}\,g\,q^{L_0 - {c\over 24}}
\bar q^{\tilde L_0 -{c\over 24}}
=\sum_{(j,\bar j)}{\chi_h^g}{}_{(j)}(q){\bar\chi^g_h}{}_{(\bar j)}(\bar q)~.
\eeq 
Each twisted sector Hilbert space ${\cal H}_h$ is decomposed  into
representations of the chiral algebra  and we introduced the corresponding
twisted chiral blocks
\beq
\label{cop5}
{\chi_h^g}{}_{(j)}(q) = \Tr_{{\mathcal{H}_h}{}_{(j)}}\,g\, q^{L_0 - {c\over 24}}~.
\eeq 
The sum over $g\in N_h$ in \eq{partfion}  projects onto $N_h$-invariant states.
Combined with their image states in sectors twisted by elements conjugate to 
$h$, states invariant under the full orbifold group $G$ are obtained.  As such,
\eq{partfion} implements the inclusion of twisted sectors and the projection
onto invariant states. In the orbifold CFT the chiral symmetry algebra is the
$G$-invariant subalgebra  $\mathcal{A}_0 = \mathcal{A}/G$ of $\mathcal{A}$. The
twisted sectors correspond to representations of $\mathcal{A}_0$ that were not
representations of $\mathcal{A}$.
\par
If the action of (subgroups of) the orbifold group $G$ admits several fixed
points  $\phi^{(I)}$, the twisted sectors $\mathcal{H}_h$ contain subsectors
$\mathcal{H}^{(I)}_h$ whenever $h$ belongs to the subgroup $G^{(I)}$ fixing
$\phi^{(I)}$. The partition function of the orbifold is still given by
\eq{partfion},  with the understanding that the terms $Z(g,h)$ receive
contributions  from each point fixed by $h$.  Below, a more explicit notation
is introduced in the expression of the partition function. The calligraphic
symbol $\mathcal{Z}(g,h)$ will denote the $h$ twisted contribution from a
\emph{single} fixed point.  The partition function then involves an explicit
sum over the fixed points:
\beq\label{partfionDTI}
Z=\frac1{|G|}\sum_{
\begin{array}{c}{g,h\in
G}\\{[g,h]=e}\end{array}}\mathcal{Z}(g,h) + \sum_{I}\,\frac1{|G^{(I)}|}
\sum_{
\begin{array}{c}{h\not=e,g\in
G^{(I)}}\\{[g,h]=e}\end{array}}\mathcal{Z}^{(I)}(g,h)~.
\eeq
The first term comes from  the fixed point implicitly associated to the
original twisted sectors  \eq{periodcondg}, whereas the second term is due to
the other fixed points. As will be pointed out in the next section, it is 
one-loop modular
invariance that necessitates the inclusion of the full set of fixed points.
\subsection{Chiral blocks and orbifold partition functions}
\label{sec:chiral}
In this section we outline in some detail how the general features of the
previous subsection are realised in the specific instance of orbifolds of free
bosons and fermions.  Technical details are collected in Appendix
\ref{app:chirblocks}.
\par
We want to model (super)string propagation on a flat $2n$-dimensional space
with a  geometrical action by some discrete group $G$. The flat space may be
either compact ($T^{2n}$) or non-compact ($\mathbb{C}^n$). The $G$-action is
implemented accordingly on the  (super)string  fields $X$ and $\psi$
corresponding to these directions. In this subsection,  we will only discuss 
the ``internal'' orbifold CFT.
Quantities referring to the remaining part of the $c=26$ or $c=15$ CFT are
suppressed.
\par
Since any $g$ and $h$ appearing together in \eq{partfion} commute, they can 
be simultaneously diagonalised.  In the sector $(g,h)$ we can therefore choose a
basis of complex fields $X^l$, $l=1,\ldots,d$ (and similarly for the $\psi$'s)
such that
\beq
\label{comp2}
g:\hskip 0.2cm X^l\mapsto \ee^{2\pi\ii\nu_{g,l}} X^l
\eeq
($\bar X^l$ transforms with opposite phases), while
the twist by $h$ imposes
\beq
\label{comp3}
X^l(\tau,\sigma+2\pi) = \ee^{2\pi\ii\nu_{h,l}}X^l(\tau,\sigma)~.
\eeq
Analogous conditions apply to the fermionic fields.
With the diagonal action \eq{comp2}, each single complex direction may be
discussed separately.  For ease of notation, we henceforth denote the
eigenvalues of $g,h$ in this direction by $\nu_g \equiv \nu, \nu_h \equiv
\nu'$.
\par
Consider the bosonic fields first. The chiral blocks $\chi^{(X)g}_h$ defined as
in \eq{cop5}, may be expressed explicitly in terms of theta-functions (see
Appendix \ref{app:chirblocks}).  Out of these, a particular subset corresponds to
the untwisted sector.  Only these blocks will appear in the open string
partition function  where strings are stretched between parallel D-branes at 
the orbifold point, so we concentrate on them in most of this paper. 
From Appendix \ref{app:chirblocks} one learns that
\bea
\label{comp6}
\widehat\chi_{e}^{(X)\,g}(q) & = & 
2\sin\pi\nu\,
{\eta(\tau)\over\theta_1(\nu|\tau)}~,
\eea
where the hat indicates the omission of the zero-modes.  The modular
transformation properties of these blocks depend crucially on the presence or
absence of  the zero-mode contribution in \eq{comp6}. An example will be given
below to illustrate this.  Rather than to deal with the general
case,   we believe it is more useful to show the peculiar role of the
zero-modes in a specific  instance. 
\par
Let us next turn to the fermions.  The relevant objects are the $v,o,s,c$
representations of the $SO(2)$ algebra . The corresponding chiral blocks in
\eq{comp5} are labelled accordingly.  The untwisted characters are
\bea
\label{untwferm2}
(\chi_v)_e^g & = 
&{\theta_3(\nu|\tau) - \theta_4(\nu|\tau)\over 2\eta(\tau)}~,
\nonumber\\
(\chi_o)_e^g & =
&{\theta_3(\nu|\tau) + \theta_4(\nu|\tau)\over 2\eta(\tau)}~,
\nonumber\\
(\chi_s)_e^g & =
&{\theta_2(\nu|\tau) - \ii\,\theta_1(\nu|\tau)\over 2\eta(\tau)}~,
\nonumber\\
(\chi_c)_e^g & =
& {\theta_2(\nu|\tau) + \ii\,\theta_1(\nu|\tau)\over 2\eta(\tau)}~.
\eea
In the last two cases, these expressions include the fermionic zero-mode
contribution (see Appendix \ref{app:chirblocks}). For explicit expressions of
$(\chi_a)_h^g$, with $a=v,o,s,c$ in the generic twisted sectors  we refer to
Appendix \ref{app:chirblocks} likewise. 
\paragraph{Modular transformations}
The modular properties of orbifold chiral blocks were considered in 
Ref.~\cite{DVVV} in the context of RCFT.  The theory of free bosons, in
particular in the non-compact case,  is not an RCFT; however the chiral blocks
are simple expressions in terms of theta-functions and as such the modular
transformations are easily derived.
\par
In the RCFT case, the generators $S:\tau\to-1/\tau$ and $T:\tau\to\tau+1$ of
the modular group of the torus are represented on the chiral blocks $\chi_h^g$ 
as follows~\cite{DVVV}:
\bea
\label{1chirmod}
\chi_h^g & \stackrel{S}{\longrightarrow} & \sigma(h|g)\,\chi_g^{h^{-1}}~;
\\
\chi_h^g & \stackrel{T}{\longrightarrow} & \ee^{-\pi\ii{c\over 12}}\,
\tau_h \,\chi_h^{hg}~.
\eea
The quantities $\sigma(h|g)$ must be symmetric in $h$ and $g$.  In the case of
orbifolds of free bosons and fermions, the modular transformations are studied
in Appendix~\ref{app:chirblocks}. They turn out to take  the form \eq{1chirmod},
with appropriate corresponding  quantities $\sigma(h,g)$ and $\tau_h$.
\par
In particular, the untwisted boson characters (without zero mode contributions)
$\widehat\chi^{(X)g}_e$ transform as follows under the $S$ modular
transformation:
\beq
\label{opmodbos}
\widehat\chi_{e}^{(X)\,g}(\tilde q) =
 2\sin\pi\nu \,\chi_{g}^{(X)\,e}(q)~,
\eeq
corresponding to 
\beq
\label{1opbosS}
\widehat\sigma(e|g) = 2\sin\pi\nu~.
\eeq
\par
For the untwisted fermions, the action of the  $S$-modular transformation is
encoded in matrices $S(h|g)$ acting on the character vector
$(\chi_a)_h^g,(a=v,o,s,c)$:
\beq
\label{1Shg}
(\chi_a)_e^g \stackrel{S}{\longrightarrow} (S_{(2)})_a^{~m}\,
(\chi_m)_g^{e}~,
\eeq
where $S_{(2)}$ is given in \eq{ven2}.
\paragraph{Bosonic zero modes in an example} 
We next wish to display the role of the bosonic zero-modes in a concrete
example.  Along the discussion, we will make a detour to comment on the
geometrical meaning of the extra factors  $2\sin\pi\nu$ present in the
S-modular transformations \eq{1opbosS} of bosonic untwisted chiral blocks. These
peculiar factors appear in the partition function of a compact orbifold, where
they count fixed points according to the Lefschetz fixed point theorem. The
very same factors $2\sin\pi\nu$ will appear in the coefficients of consistent 
boundary states for fractional D-branes sitting at the orbifold singularity.
\par
Given its mainly illustrative purposes, the discussion can be restricted to the
simple case of one complex dimensional abelian orbifolds; that is, to a 
$\mathbb{Z}_N$ point group action.
\par
Consider the non-compact case first. Denote the $\mathbb{Z}_N$ generator by
$g$.  According to \eq{partfion}, the orbifold partition function contains
$Z_{\mathrm{flat}}$,  all the contributions $Z_{\mathrm{n.c.}}(e,g^\a)$ needed
to project onto  $\mathbb{Z}_N$ invariant states in the untwisted sector.  The
zero modes (the momenta  $\mathbf{k}$) multiply the non-zero mode contribution
\eq{opmodbos} as follows: 
\bea
\label{sera2}
Z_{\mathrm{n.c.}}(e,g^\a) & = & 
\int d^2\mathbf{k}\bra{\mathbf{k}}g^\a q^{\frac{\alpha'\mathbf{k}^2}2} 
\ket{\mathbf{k}}|\widehat\chi_{e}^{(X)\,g^\a}|^2
= \int d^2\mathbf{k}\delta^2\left((1 - g^\a)\mathbf{k}\right)
q^{\frac{\alpha'\mathbf{k}^2}2}|\widehat\chi_{e}^{(X)\,g^\a}|^2
\nonumber\\
& = & \mathrm{det}(1 - g^\a)^{-1}\,|\widehat\chi_{e}^{(X)\,g^\a}|^2
=  {|\widehat\chi_{e}^{(X)\,g^\a}|^2\over 4\sin^2\pi{\a\over N}} =
\left|{\eta(\tau)\over \theta_1({\a\over N}|\tau)}\right|^2~.
\eea
Orthogonality of momentum eigenstates was used here. The eigenvalues of the
$g^\a$  on $\mathbf{k}$  are obviously $\ee^{\pm 2\pi\ii\nu}$, with
$\nu=\a/N$.  As such, the trigonometric factors $2\sin\pi\nu$ are seen to be
absorbed by the momentum integration.
\par
In the twisted sectors there is no momentum, so that for $\b\not=0$ we simply find
\beq
\label{untw}
\mathcal{Z}(g^\b,g^\a)\equiv Z_{\mathrm{n.c.}}(g^\b,g^\a) =
|\chi^{(X)g^\a}_{g^\b}|^2
\eeq 
(see Appendix \ref{app:chirblocks} for explicit expressions of the twisted
chiral blocks  $\chi^{(X)g^\a}_{g^\b}$). 
\par
The modular gen\-er\-at\-ors $S$ and $T$ are thus seen to act on the
$Z_{\mathrm{n.c.}}(g^\b,g^\a)$, for all $\b,\a$, by an appropriate exchange of
the boundary cond\-it\-i\-ons (see \eq{1chirmod}). Therefore, the partition
function corresponding to \eq{partfion},
\beq
\label{forgot1}
Z_{\mathrm{n.c.}} = 
{1\over N} \sum_{\b,\a=0}^{N-1} Z_{\mathrm{n.c.}}(g^\b,g^\a)
=  {1\over N} \left\{Z_{\mathrm{flat}} + \sum_{\a=1}^{N-1} 
Z_{\mathrm{n.c.}}(e,g^\a)\right\} + 
\mathcal{Z}^{\mathbb{Z}_N}_{\mathrm{twisted}}~,
\eeq
is modular invariant. The notation used here is such that
$\mathcal{Z}^{\mathbb{Z}_N}_{\mathrm{twisted}}$ is the contribution of twisted 
sectors from a single $\mathbb{Z}_N$ fixed point: 
\begin{equation}
\label{lun2}
\mathcal{Z}^{\mathbb{Z}_N}_\mathrm{twisted} = {1\over N} 
\sum_{\b=0}^{N-1}\sum_{\a=1}^{N-1} |\chi^{g^\b}_{g^\a}|^2~,
\end{equation}
where $g$ generates $\mathbb{Z}_N$.
\par
Next consider the compact case, where  $\mathbb{C}$ is compactified
on a  $\mathbb{Z}_N$ invariant lattice $\Lambda$. Geometrically, the only
consistent choices, giving rise to a quotient space with isolated orbifold
singularities, are $N=2,3,4,6$. 
\par
Momenta take values in the dual lattice. Furthermore, there are now closed
string winding states. The unprojected, untwisted partition function 
$Z_{\mathrm{torus}}$ that is now obtained is again modular invariant by itself.
The untwisted contribution performing the projection onto invariant states is
now found to be
\bea 
\nonumber
\mathcal{Z}(e,g^\a) &=& \sum_{\mathbf{n},\mathbf{w}} 
\bra{\mathbf{n};\mathbf{w}}g^\a\, q^{\frac{\alpha'}2{\mathbf{n}^2\over R^2} + 
 {(\mathbf{w}R)^2\over{2 \alpha'}}} \ket{\mathbf{n};\mathbf{w}}\,
|\widehat\chi_{e}^{(X)\,g^\a}|^2 = |\widehat\chi_{e}^{(X)\,g^\a}|^2~, \\
\label{sera1}
&=& 4\sin^2{\pi \a\over N}\ Z_{\mathrm{n.c.}}(e,g^\a)\ ,
\eea
since only $\ket{\mathbf{n}=0;\mathbf{w}=0}$ is invariant under the  action of
$g$. Contrary to the non-compact case, the $S$ modular transform of the
untwisted terms is that of the chiral blocks $\chi_{e}^{(X)\,g^\a}$ in
\eq{opmodbos}. More precisely, we have
\beq
\label{vc2}
\mathcal{Z}(e,g^\a) \stackrel{S}{\longrightarrow} 
4 \sin^2\left( {\pi \a\over N}\right)\,\, \mathcal{Z}(g^{\a},e)~.
\eeq
Adding all sectors with weight one like in \eq{forgot1} would no longer yield a
modular invariant. \eq{vc2} requires that  the terms  $(1/N)\sum_{\b=1}^{N-1}
4\sin^2(\pi \b/N)$ $\, \mathcal{Z}(g^\b,e)$ be added to the un\-twist\-ed
pro\-ject\-ed con\-trib\-u\-ti\-on $\frac 1N (Z_{\mathrm{torus}} +
\sum_{\a=1}^{N-1} \mathcal{Z}(e,g^\a))$.  To ensure full modular invariance,
the remaining sectors must be included with appropriate weigths determined by
these.
\par
It is only in the cases $N=2,3,4,6$ that the factors $N_\b = 4 \sin^2 (\pi
\b/N)$ are all integers. These are precisely the cases where a geometrical
interpretation of the quotient space is possible. More precisely, $N_\b$ counts
the number of points on $T^2$ fixed under the action of the element $g^\b$ of
the orbifold group; this is basically the content of the Lefschetz fixed-point
theorem. As such, the modular invariant partition function fixed in terms of
the factors $N_\a$ in \eq{vc2} incorporates the contributions of twisted
sectors from all the points fixed under (subgroups of) the orbifold group. In
other words, it has exactly the structure of \eq{partfionDTI}. 
\FIGURE{\epsfig{file=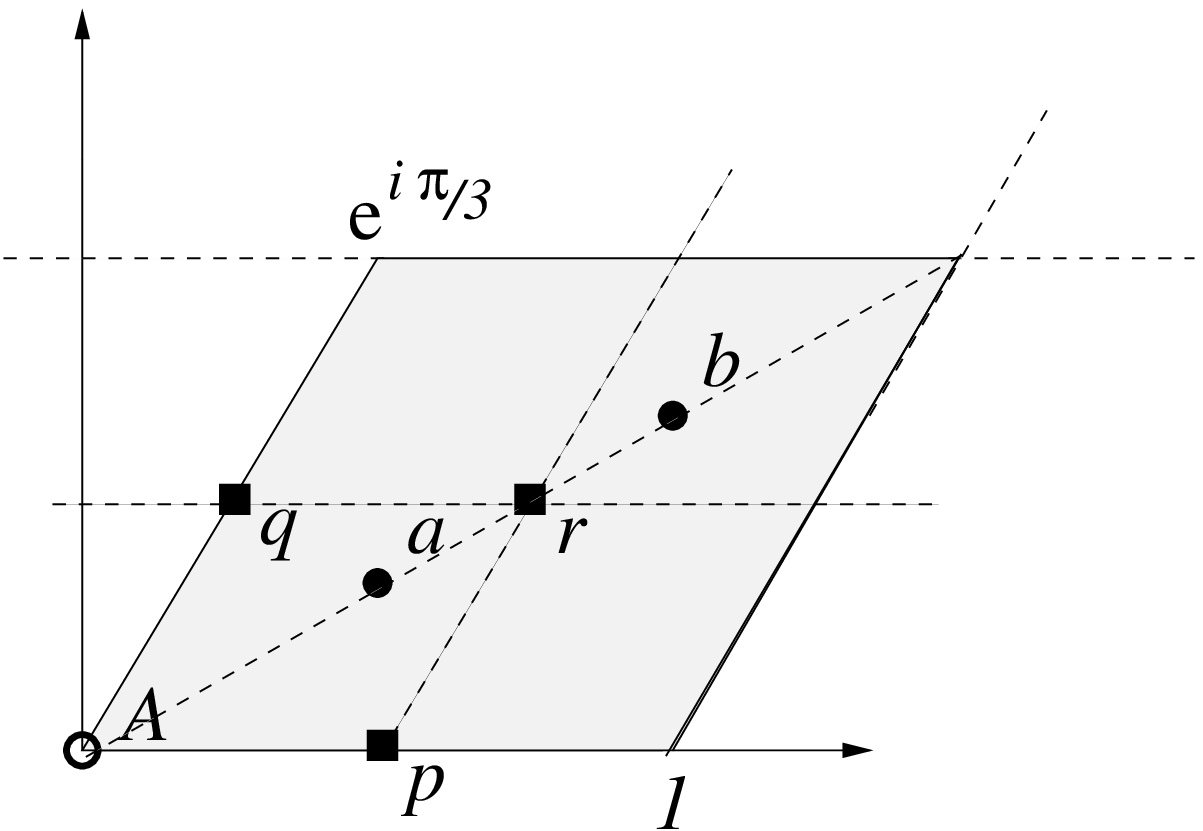,width=7cm}
\caption{Fixed points for $T^2/\mathbb{Z}_6$. See the text for details.}
\label{fig2}}
\par
Let us illustrate this in the concrete example of $T^2/\mathbb{Z}_6$. There are
several fixed points now. There is one  $\mathbb{Z}_6$-fixed point $A$ (fixed
by the generator $g$) in Fig.~\ref{fig2}. Further, there are  two
$\mathbb{Z}_3$-fixed points $a$ and $b$ in Fig.~\ref{fig2}, (fixed by $g^2$). 
These points are combined in a \emph{single} orbit under the  $\mathbb{Z}_6$
action of $g$. Finally, there are 3 $\mathbb{Z}_2$-fixed points ($p$,$q$ and
$r$), fixed by $g^3$; they form a \emph{single} orbit under  the $\mathbb{Z}_6$
action.  The modular invariant partition function is
\bea
\label{sc1}
Z & = & {1\over 6}\Bigl\{ Z_{\mathrm{torus}} + \sum_{\a=1}^5 
\mathcal{Z}(e,g^\a) + N_1  \sum_{\a=0}^5 \mathcal{Z}(g,g^\a) + 
N_2 \sum_{\a=0}^2 \mathcal{Z}(g^2,g^{2\a})
\Bigr.
\nonumber\\
& + & N_3 \sum_{\a=0}^1 \mathcal{Z}(g^3,g^{3\a}) 
+ N_4 \sum_{\a=0}^3 \mathcal{Z}(g^4,g^{2\a})
+ N_5 \sum_{\a=0}^5 \mathcal{Z}(g^5,g^\a)
\nonumber\\
& + & N_1  \Bigl.\sum_{\a=1,3,5} \mathcal{Z}(g^2,g^\a) + 
N_1 \sum_{\a=1,2,4,5} \mathcal{Z}(g^3,g^\a)
+ N_1 \sum_{\a=1,3,5} \mathcal{Z}(g^4,g^\a) \Bigr\}~, 
\eea
with $N_1=N_5 = 4 \sin^2(\pi/6) = 1$, and analogously $N_2=N_4=3$ and
$N_3=4$.  In this expression, all possible twisted contributions at points
fixed by elements $g^\b$ appear correctly counted with the fixed point multiplicities
$N_\b=N_{6-\b}$. This partition function is rewritten more suggestively in the form of
\eq{partfion}:
\beq
\label{sc2}
Z =  {1\over 6} \Bigl\{Z_{\mathrm{torus}} + 
\sum_{\a=1}^5 \mathcal{Z}(e,g^\a)\Bigr\}
+ \mathcal{Z}^{\mathbb{Z}_6}_{\mathrm{twisted}}
+ \mathcal{Z}^{\mathbb{Z}_3}_{\mathrm{twisted}}
+ \mathcal{Z}^{\mathbb{Z}_2}_{\mathrm{twisted}}~,
\eeq
using the notation introduced in \eq{lun2}. The $\mathbb{Z}_3$ and 
$\mathbb{Z}_2$ twisted contributions appear only once, as we are now
identifying  the fixed points belonging to the same orbits.
\subsection{Boundary states for fractional branes}
\label{sec:fractional}
Here and below, we will restrict our attention to D-branes that are pointlike along the
orbifold directions. The orbifold space is taken to be $\mathbb{C}^n/\Gamma$; following \cite{
Douglas:1996sw} a natural starting point are $\Gamma$-invariant configurations of D-branes on
the covering space $\mathbb{C}^n$. 
\FIGURE{
\epsfig{file=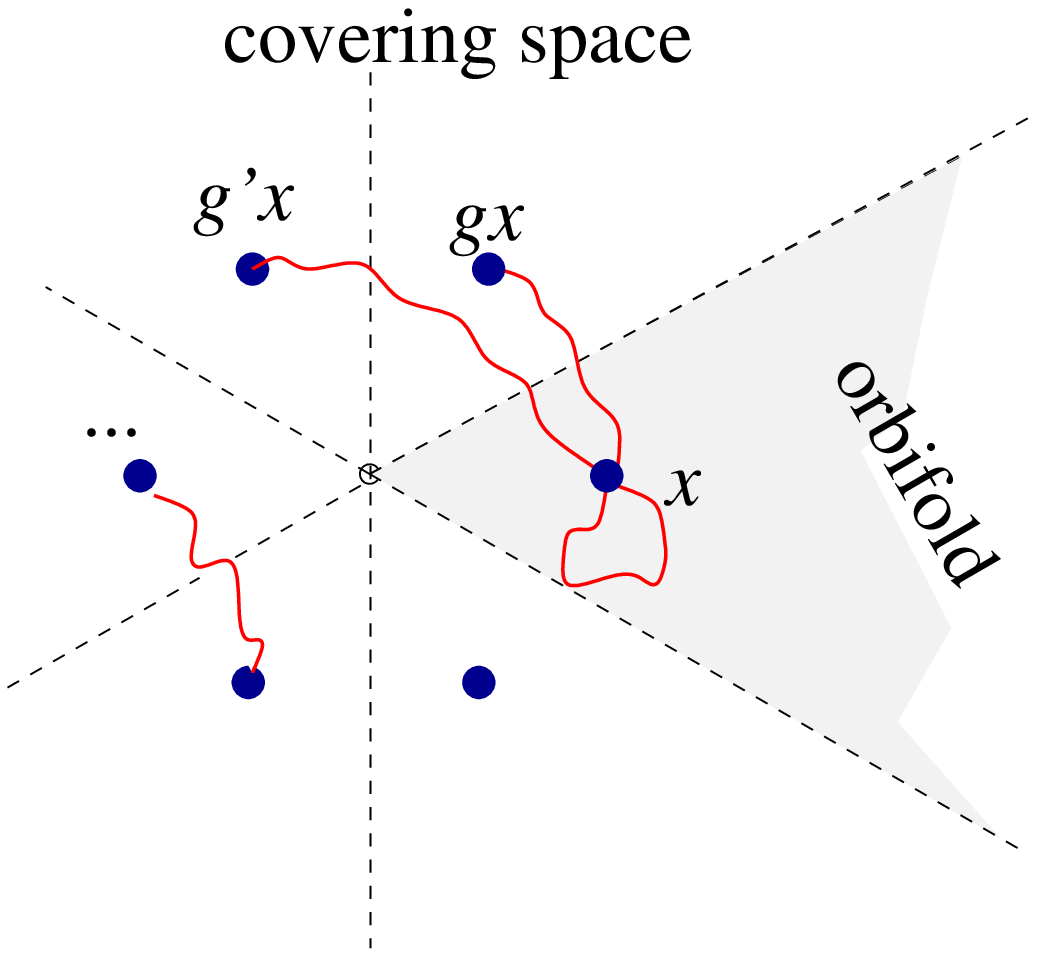,width=6cm}
\caption{A $\Gamma$-invariant configuration of D-branes (blue dots) in the
covering spaces. Open strings carry Chan-paton indices in the regular
representation $\mathcal{R}$.}\label{fig0a}}
As such the brane and its images naturally give rise to the
regular representation\footnote{%
The regular representation is the representation of dimension $|\Gamma|$
obtained regarding the group products as linear operations on a  vector space
spanned by the group elements themselves. Namely, the matrix $\mathcal{R}(g)$
representing $g$ has matrix elements $[\mathcal{R}(g)]_{g_1 g_2} =
\delta_{g_2,g\cdot g_1}$.} 
$\mathcal{R}$ on the Chan-Paton factors. Correspondingly, these D-branes have
been called regular branes. 
A particularly well-studied example is given by
regular D-branes on $\mathbb{C}^2/\Gamma$, where the low-energy world-volume
theory yields a supersymmetric $\sigma$-model on the ALE-space resolving the
singularity.%
\footnote{This has been rephrased as follows
~\cite{Douglas:1996sw,Johnson:1997py}: the matter fields of the world--volume
theory give a physical realisation of the mathematical hyperk\" ahler quotient
construction of ALE spaces, originally due to Kronheimer \cite{kronheimer}.}
%
\par
However, the regular representation of a discrete group $\Gamma$ is not
irreducible. It decomposes as
\beq
\label{mm1}
\mathcal{R} = \bigoplus_I d_I\, \mathcal{D}^I~, 
\eeq
in terms of the irreducible representations $\mathcal{D}^I$ of $\Gamma$,  whose
dimensions we denote by $d_I$. Thus one may wonder whether there exists a 
more ``elementary'' set of D-branes such that the open strings  attached to the latter
carry Chan-Paton indices transforming in an irreducible representation.  This
turns out to be the case, and such D-branes have been called  \emph{fractional}
D-branes~\cite{Polchinski:1997ry,Douglas:1997xg,
Douglas:1997de,Diaconescu:1998br} (see also  \cite{Johnson:2000ch} for a
review). 
\par
A fractional D$p$-brane is a BPS object. It carries only a fraction of the
charge with respect to the untwisted RR $(p+1)$--form of a usual D$p$-brane
but, contrarily to a usual brane, it is charged with respect to some twisted
RR $(p+1)$--form~\cite{Douglas:1997xg}. It is stuck at the orbifold fixed point (where all twisted
fields sit): indeed, to place it elsewhere, we should associate it to an
invariant configuration in in the covering space. However, as already
discussed (see Figure \ref{fig0a}), such an invariant configuration
corresponds to the regular representation. 
\FIGURE{
\epsfig{file=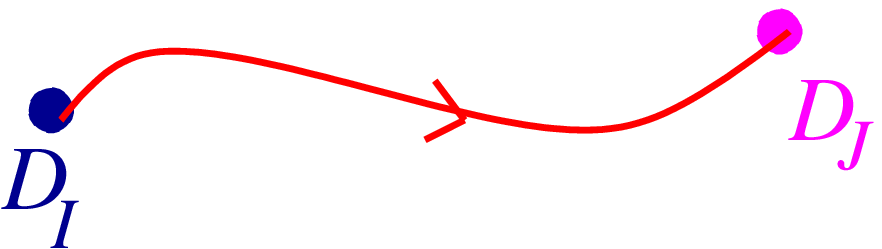,width=5cm}
\caption{Open strings between fractional branes of types $I$
and $J$ carry Chan-Paton labels in the irreducible representations 
$\bar\mathcal{D}_I$ and $\mathcal{D}_J$.}\label{fig0b}}
\par
Fractional branes can be geometrically interpreted as higher-dimensional
branes wrapped  on exceptional cycles of the resolved
space~\cite{Douglas:1997xg,Diaconescu:1998br}. The exceptional cycles collapse
in the orbifold limit, but a non-zero flux of $B$ on them persists  
\cite{Aspinwall:1995zi}, leaving us with lower-dimensional
fractional branes with non-zero mass. The determination of the spectrum of
D-branes in the theory should thus be linked with the determination  of the
homology of the (resolved) orbifold space or, more precisely, to its
homological K-theory  \cite{Garcia-Compean:1999rg,Diaconescu:2000dt,
Gaberdiel:2000ch,Aspinwall:2000xs}.   
\par
As already mentioned, we concentrate solely on the fractional branes. Our aim is to 
derive the corresponding consistent boundary states by adapting Cardy's
construction\footnote{%
The boundary states for fractional branes have already  been considered in the
literature \cite{Diaconescu:2000dt,Takayanagi:2000rd,Roose:2000ay}.  Their
derivation via Cardy's procedure, however,  will allow us to clarify several
fine points.}. 
This procedure applies equally well to abelian and non-abelian orbifolds, and
it links very explicitly  (for instance in the case of $\mathbb{C}^2/\Gamma$)
the boundary states that  represent the fractional $p$-branes  at the orbifold
point with  the geometrical picture of such branes as higher-dimensional 
branes wrapped on vanishing cycles.  
\par
\FIGURE{
\epsfig{file=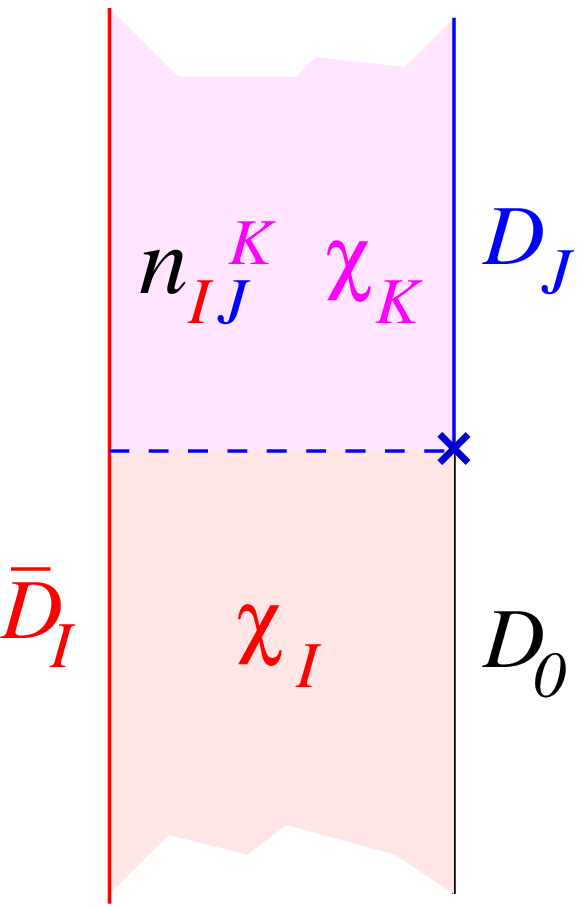,width=4cm}
\caption{The boundary conditions corresponding to fractional branes
fuse in correspondence to the algebra of irreducible representations of 
$\Gamma$.}\label{fig2a}}
We concentrate first on describing the boundary states at the level of the 
CFT that describes the $\mathbb{C}^n/\Gamma$ orbifold space.  In terms of
these it is straightforward to write the boundary states of the complete
string theory under consideration, be it a bosonic or a supersymmetric string
theory.
\par 
The cylinder amplitudes with fixed boundary conditions of type $I$ and $J$ are
described, from the point of view of open strings, as  $\Gamma$-projected
one-loop traces
\beq
\label{mk1}
Z_{IJ}(q) = 
{1\over |\Gamma|} \sum_{g\in \Gamma} \Tr_{IJ}(\hat g\, q^{L_0 - {c\over
24}})~,
\eeq     
where $\hat g$ acts not only on the string fields $X^\mu$ and $\psi^\mu$, but
also on the Chan-Paton labels at the two open string endpoints, transforming
respectively in the representation $\bar\mathcal{D}_I$ and $\mathcal{D}_J$.
The amplitude $Z_{IJ}$ is basically constructed out of the untwisted chiral
blocks $\chi_e^g(q)$ discussed in Sec.~\ref{sec:chiral}. Indeed,
performing the Chan-Paton part of the traces, we have
\bea
\label{mk2}
 Z_{IJ}(q) & = & {1\over |\Gamma|} \sum_{g\in \Gamma}\, 
 \tr_{\bar\mathcal{D}_I}(g)\,
 \tr_{\mathcal{D}_J}(g)\, \Tr\,(g\, q^{L_0 - {c\over
24}})\nonumber\\
& = &  {1\over |\Gamma|} \sum_\alpha\, n_\alpha\,(\rho_I^\alpha)^*
\rho_J^\alpha\, \chi_e^{g^{(\alpha)}}(q)~.
\eea 
Here and below the index $\alpha$ will label the conjugacy classes $\mathcal{C}^\alpha$
(containing $n_\alpha$ elements) of $\Gamma$ and the group element 
$g^{(\alpha)}$ is a representative of $\mathcal{C}^\alpha$; moreover, 
$\rho_J^\alpha$ denotes the character matrix,  namely $\rho_J^\alpha =
\tr_{\mathcal{D}_J}(g^{(\alpha)})$.
\par
The consistent amplitudes $Z_{IJ}$ can be expanded with integer multiplicities
on a basis of chiral blocks, $\chi_I(q)$: 
\beq
\label{mk3}
Z_{IJ}(q) = \sum_{K} n_{IJ}^K\, \chi_K(q)~.
\eeq   
The chiral blocks $\chi_I$ are amplitudes in which only the Chan-Paton
representation $\mathcal{D}_I$ runs in the loop. They correspond to amplitudes
$Z_{0I}$, where at one endpoint the Chan-Paton labels transform in the trivial
identical representation $\mathcal{D}_0$. It follows from \eq{mk2} that these
blocks are simply the discrete Fourier transforms of the chiral blocks
$\chi_e^g$:
\beq
\label{mk4}
\chi_I(q) = {1\over|\Gamma|} \sum_\alpha\, n_\alpha\,
\rho_I^\alpha\, \chi_e^{g^{(\alpha)}}(q)~.
\eeq 
\FIGURE{
\epsfig{file=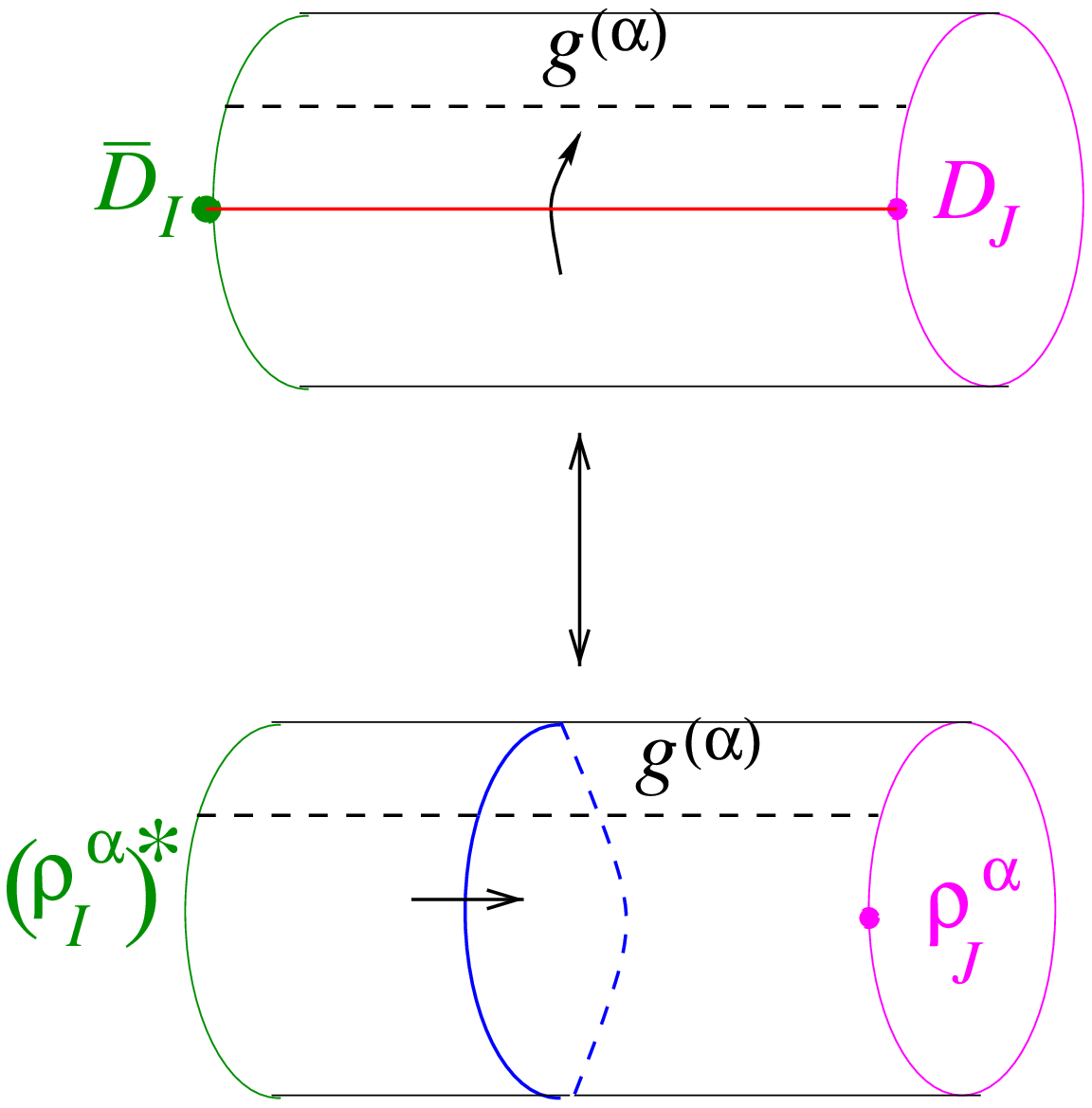,width=4.5cm}
\caption{The open string one loop trace  with b.c.s of type $I$ and $J$ 
with the insertion of an element $g^{(\a)}$ (top) corresponds upon 
$S$-modular and discrete Fourier transform to the propagation 
in the tree channel of a closed string twisted by $g^{(\a)}$ (bottom).
$\rho$ denotes the character matrix.} 
\label{fig2b}}
%
%
\par
To carry out Cardy's construction, we need the $S$-modular transformation
properties of the open string basis $\chi_I(q)$ of chiral blocks. Via \eq{mk4},
these are determined by the transformations of the untwisted chiral blocks
$\chi_e^g$.
\paragraph{Bosonic theory}
Consider first the purely bosonic case. Recall that we consider open strings 
with Dirichlet boundary conditions in all of the orbifold directions, so that
there is no momentum zero-mode. The $S$ modular transformation of the relevant
chiral blocks $\chi_e^g = \hat\chi^{(X)g}_h$ of \eq{comp6} reads 
\beq
\label{mk6}
\chi_e^g\stackrel{S}{\to} \sigma(e,g)\,\chi_g^e~,
\eeq
where  the factors $\sigma(e,g)$ are as follows (see \eq{opmodbos}):
\bea
\label{mk7}
\sigma(e,e) & = & 1~, \nonumber\\
\sigma(e,g) & = & \prod_{l=1}^n (2\sin\pi\nu_{g,l})~,\hskip 0.8cm g\not= e~,
\eea
if we are considering a $\mathbb{C}^n/\Gamma$ orbifold.
\par
From \eq{mk4} and \eq{mk6} it follows that the $\chi_I$ blocks, when 
transformed to the tree channel, are expressed them in terms of twisted chiral
blocks. Twisted blocks are labeled by the conjugacy classes of $\Gamma$;  it
is convenient to introduce the simplified notation $\chi_\a(\tilde q)$ for a
basis of the tree-channel blocks:
\beq
\label{mk8}
\chi_\a(\tilde q) = \chi_{g^{(\a)}}^e(\tilde q)~,
\hskip 0.8cm g^{(\a)}\in\mathcal{C}^\a~.
\eeq
We can then write the modular transformation as
\beq
\label{mk9}
\chi_I(q) = \sum_\a \mathcal{S}_I^\alpha\, \chi_\a(\tilde q)~,
\eeq
with
\beq
\label{mk8bis}
\mathcal{S}_I^\a = {n_\alpha\, \rho_I^\alpha\over|\Gamma|} 
\, \sigma(e,g^{(\a)})~.
\eeq
Notice that the entries of  $\mathcal{S}_0^\a$, namely  
$\sigma(e,g^{(\a)})\,n_\alpha/|\Gamma|$, are all positive, as appropriate to
the distinguished role of the identity representation in Cardy's procedure.  
The fusion algebra that follows  from applying Verlinde's formula to the matrix
$\mathcal{S}$,
\beq
\label{mk9bis}
n_{IJ}^K = \sum_\a {\mathcal{S}_I^\a\mathcal{S}_J^\a (\mathcal{S}^{-1})_\a^K
\over \mathcal{S}_0^\a}~,
\eeq
corresponds simply\footnote{%
The inverse of the matrix $\mathcal{S}$ of \eq{mk8} is given by
$(\mathcal{S}^{-1})_\a^J = (\rho_J^\a)^*/\sigma(e,g^{(\a)})$, as it follows
from the orthogonality relations  of the characters.  Then in the Verlinde
formula \eq{mk9} all factors $\sigma(e,g^{(\a)})$ cancel and one gets 
\beq
n_{IJ}^K = {1\over |\Gamma|}\sum_\a n_a\, \rho_I^\a\,\rho_J^\a\,(\rho_k^a)^*~,
\eeq
which is also the expression that follows from \eq{mk10}, using again the
orthogonality properties of characters.}
to the algebra of the irreducible representations of $\Gamma$:
\beq
\label{mk10}
\mathcal{D}_I\otimes\mathcal{D}_I = \sum_K n_{IJ}^K \, \mathcal{D}_K~.
\eeq
\par
In terms of this matrix $\mathcal{S}$ we can now proceed with 
Cardy's construction of consistent boundary states, as in \secn{sec:cardysol}.
\par
``Ishibashi states'' $\dket{g^{(\a)}}$ solve the analogue of the gluing
condition \eq{genbndcnd} (of Dirichlet type) in a sector twisted by $g^{(\a)}$ 
(see \cite{Billo:1999nf}):
\beq
\label{mk11}
\dket{g^{(\a)}} = \exp \left( \sum_{l=1}^n \left[
\sum_{\kappa_l} {\bar\a^{l}_{-\kappa_l} \tilde\a^l_{-\kappa_l}\over \kappa_l} 
+ \sum_{\bar\kappa_l} {\a^{l}_{-\bar\kappa_l} 
\tilde{\bar\a}^l_{-\bar\kappa_l}\over 
\bar\kappa_l}\right]\right)\, \ket{0,g^{(\a)}}~.
\eeq
The modings of the oscillators $\alpha^l_{\kappa_i}$ follows from the twisted
identifications  \eq{comp3}, and were implicitly used in the computations of
the twisted chiral blocks: we have $\kappa_l\in\mathbb{Z}+\nu_{\a,l}$, while  
$\bar\kappa_l\in\mathbb{Z}-\nu_{\a,l}$, where $\exp(2\pi\ii\nu_{\a,l})$ is the
eigenvalue of $g^{(\a)}\in \mathcal{C}^\a$ on the complex field $X^l$. By
$\ket{0,g^{(\a)}}$ we have denoted the vacuum in the sector twisted by
$g^{(\a)}$.  All the directions in the orbifold space are uncompactified 
Dirichlet directions; if in the $l$-th complex direction the sector defined by 
$g^{(\a)}$ appears as untwisted (\emph{i.e.}, if $\nu_{\a,l}=0$), there is a   
zero-mode,  and the boundary state includes a  zero-mode part of the type
$\delta^{2}(\hat X^l) \ket{k^l=0}$, as discussed in \secn{bosbsflat}. 
\par
In particular, the untwisted part of the boundary state is the only one that
emits closed strings that carry momentum  in all of the orbifold directions; as
we shall see, also the normalization of the untwisted component of the 
boundary state in the full string theory differs from the one of the twisted
components. 
\par
In fact, the ``Ishibashi states'' $\dket{g^{(\a)}}$ are not quite what we want:
for one thing, in
the case of a non-abelian $\Gamma$
they are not invariant under the orbifold projection. Rather, they mix with 
``Ishibashi states'' in sectors twisted by conjugate group
elements. The invariant Ishibashi state associated to a conjugacy class
$\mathcal{C}^\a$ reads
\beq\label{invIshi}
\dket{\a}=\frac1{\sqrt{n_\alpha}}\sum_{g^{(\a)}\in\mathcal{C}^\a}
\dket{g^{(\a)}}~.
\eeq
The states $\dket{\a}$ are orthogonal, satisfying:
\beq
\label{mk12}
\dbra{\a} \tilde q^{\frac 1 2 ({L_0 +
\tilde L_0 -\frac c {12}})} \dket{\b} = \delta_{\a\b}\,\chi_\a(\tilde q)~.
\eeq
\par
The consistent boundary state associated to a boundary condition of type
$I$ (\emph{i.e.}, with Chan-Paton labels in the representation $\mathcal{D}_I$)
is then obtained via Cardy's formula \eq{coeffS} and can be written as
\bea
\label{mk13}
\ket{I} & = & 
{1\over\sqrt{|\Gamma|}} \sum_\a
\sqrt{n_\a\,\sigma(e,g^{(\a)})}\,\rho_I^\a\,\dket{\a}
\nonumber\\
& = & \sum_\a \psi_I^\a\, \sqrt{\sigma(e,g^{(\a)})}\, \dket{\a}~,
\eea
where we introduced the quantities
\beq
\label{mk13a}
\psi_I^\a = \sqrt{n_\a\over |\Gamma|}\,\rho_I^\a~.
\eeq
In terms of the boundary states $\ket I$, the open-string one-loop
amplitude $Z_{IJ}(q)$, transformed to the tree channel, is simply
\beq
\label{mk13b}
Z_{IJ}(\tilde q)=\bra{I} \tilde q^{\frac 1 2 ({L_0 +
\tilde L_0 -\frac c {12}})} \ket{J}~.
\eeq
\paragraph{(Super)string theory}
All the expressions obtained so far must be supplemented with the  contributions
from the directions transverse to the orbifold space. Furthermore, the
appropriate normalisations must be taken into account in order to get the 
full-fledged physical string boundary states that give the closed string tree-channel
description of the cylinder open string  \emph{amplitude}. The only subtlety
in completing this programme is common to both the bosonic and the 
superstring, and resides in a
proper treatment of \emph{closed} string bosonic zero-modes (momenta) along
orbifold directions.
\par
The one-loop open string amplitude $\mathcal{Z}_{IJ}$ corresponding to 
boundary conditions  $I$ and $J$ in the orbifold CFT and with $p+1$
world-volume directions (transverse to the orbifold) is given by
\eqn{1stringbs1} with the flat cylinder partition function $Z_{1,2}^{(d)}(q)$
replaced by  $Z^{(d-2n)}_{1,2}(q)\, Z_{IJ}(q)$, where $Z_{IJ}$ was given in
\eq{mk1}. In the case of the superstring, the fermionic sectors, discussed in
\secn{sec:fermbsflat} for the flat space part, and in next paragraph for  the
orbifold CFT, must be taken into account appropriately. 
\par
Discrete Fourier and modular transformations turn
the open string amplitude $\mathcal{Z}_{IJ}$ into the amplitude for closed string
propagation between boundary states $\dket{B,\a}$, exactly  as in
\eq{1stringbs2}. 
Apart from the orbifold Ishibashi states $\dket{\a}$, the full Ishibashi states
$\dket{B,\a}$ also contain  flat space components of \secn{bosbsflat} (and
\secn{sec:fermbsflat}) accounting for the transverse directions. Moreover, they
have to be properly normalised so as  to ensure the  equality of the amplitude
in the two channels. These normalisations are readily  obtained following the
discussion in \cite{DiVecchia:1999rh} that leads to the usual normalisation
$N_p$ of \eq{1stringbs3}. One sees that  the more general result in our
situation is 
\beq
\label{stringbs3}
N_p^{(\a)} = \frac{\sqrt{\pi}}{2}\, 2^{\frac{10 - d^{(\a)}}{4}}\,
(2\pi\sqrt{\a'})^{\frac{d^{(\a)} -4}{2} -p}~,
\eeq
where $d^{(\a)}$ is the number of Neumann directions ($p+1$) plus the number of
Dirichlet directions in which the  $\delta(\hat x)\ket{k=0}$  zero-mode part 
(see the final remarks in Appendix \ref{bosbsflat}) 
is present. Thus, in the untwisted case
we have $d^{(0)}=d$ and we get  \eq{1stringbs3}, while in the twisted
sectors\footnote{%
Assuming for simplicity that in the twisted sector no zero-modes are present 
in any of the complex directions along the orbifold, see the discussion after
\eq{mk11}.}
$\a\not=0$ we have $d^{(\a)}=d-2n$. The charges in the twisted
sectors are thus given by the analogues of 
\eq{RRcharge}, namely by 
\beq\label{twisttmu}
\mu^ {(\a)}_p = 2\sqrt{2}\, 2^{-{n\over 2}}\, N_p^{(\a)} =
\sqrt{2\pi}\, (2\pi\sqrt{\a'})^{3-n-p} 
\eeq
for $d=10$. Eq. (\ref{stringbs3}) agrees with the results of
\cite{Takayanagi:2000rd}.
\par
The \emph{untwisted} fields (such as the graviton and dilaton and the untwisted
RR forms) emitted by the D$p$-brane at the orbifold fixed point propagate in
all directions, including those along the orbifold. 
To have canonically normalised bulk kinetic terms for these fields, 
which depend on all the coordinates of $\mathbb{R}^{1,d-2n-1}\times 
\mathbb{C}^n/\Gamma$, we define them as $\sqrt{|\Gamma|}$ times the
fields that are canonically normalised on the covering space
$\mathbb{R}^{1,d-1}$. As a consequence, the tension and charge normalisations
$T_p^{(\Gamma)}$ and $\mu_p^{(\Gamma)}$ with respect to the canonically 
normalised fields in $\mathbb{R}^{1,d-2n-1}\times  \mathbb{C}^n/\Gamma$ are 
related to the flat space expressions given in \eq{1stringbs3} and
\eq{RRcharge}  respectively; in particular,
\beq\label{tmugamma}
T_p^{(\Gamma)} = \frac{T_p}{\sqrt{|\Gamma|}}~, \hskip 1cm
\mu_p^{(\Gamma)} = \frac{\mu_p}{\sqrt{|\Gamma|}}~.
\eeq
\par 
No such redefinition is required for the twisted fields, since these do not
propagate in the orbifold.
Notice that a further factor of $1/\sqrt{|\Gamma|}$ is incorporated in the 
orbifold states $\ket{I}$ (both twisted and untwisted) via the definition
\eq{mk13a} of the $\psi^\a_I$ coefficients. 
\paragraph{Fermionic sectors}
In the superstring case, the transformation of the fermionic parts adds no
complication beyond those already discussed for flat space. The relevant chiral
blocks are now products of the bosonic and fermionic ones given in
\secn{sec:chiral} and Appendix \ref{app:chirblocks}. Let us denote them by
\beq
\label{mk14}
(\chi_a)_h^g = \hat\chi^{(X)g}_h\,(\chi^{(\psi)}_a)_h^g~,
\eeq
with $a=v,o,s,c$. Then the $S$-modular transformation of an untwisted chiral
block reads
\beq
\label{mk15}
(\chi_a)_e^g\stackrel{S}{\to} \sigma(e,g)\, \sum_m (S_{(2n)})_a^m (\chi_m)_g^e~,
\eeq
where $S_{(2n)}$ is the modular matrix for the fermionic characters of
$\mathrm{SO}(2n)$ (if we are considering a $\mathbb{C}^n/\Gamma$ orbifold), 
given in \eq{ven2}, and the factors $\sigma(e,g)$ are again those of \eq{mk7}. 
We see that the modular transformation consists in a direct product of the 
transformation acting on the $\mathrm{SO}(2n)$ labels $v,o,s,c$ and on the
$\Gamma$-labels. Thus one can straightforwardly proceed to define a basis of
open-string chiral blocks $\chi_{I,a}(q)$ (with $a=v,o,s,c$) and one of
tree-level ones $\chi_{\a,m}(q')$ (with $m=v,o,s,c$), related by the modular
transformation
\beq
\label{mk16}
\chi_{I,a}(q) = \sum_\a \sum_m \mathcal{S}_I^\a\,(S_{(2n)})_a^m\,
\chi_{\a,m}(q')~.
\eeq  
As far as boundary states are concerned, one starts, analogously to the flat
space case, with states
\beq
\label{mk17} 
\dket{\a;\sigma;\eta} = \dket{\a}\,\dket{\a;\sigma;\eta}_\psi~,
\eeq
with $\sigma=\mathrm{NS,R}$ and $\eta=\pm$.
In \eq{mk17} the bosonic states $\dket{\a}$ of \eq{mk11} are 
multiplied by fermionic ones $\dket{\a;\sigma;\eta}_\psi$,
which solve the overlap condition for the (complex) fermionic oscillators:
\beq
\label{mk18}
\psi^l_{r_l} = \ii\eta\,\tilde\psi^l_{-r_l}~,
\eeq
($l=1,\ldots n$), where the modings $r_l$ are dictated by the $g^{(\a)}$ 
twist and the R or NS sector: $r_l\in \mathbb{Z}+\nu_{\a,l}$ in the R sector,
and an extra shift of 1/2 in the NS sector.\footnote{%
The explicit solution to the condition \eq{mk18} is in terms of a
coherent-state--like expression analogous to the bosonic one, \eq{mk11}, see,
\emph{e.g.}, \cite{Billo:1999nf}. Moreover, when fermionic  zero-modes are
present, extra contributions analogous to the ones in \eq{Rzeromodes} must be
taken into account.}
Then a proper orthogonal basis of Ishibashi states 
$\dket{\a,m}$ ($m=v,o,s,c$) is defined by taking combinations as in 
Eqs.~(\ref{NS+}-\ref{R-}). Finally, consistent Cardy states (for instance, in type~0
theory) $\ket{I,a}$, with $a=v,o,s,c$, are given by
\beq
\label{mk19}
\ket{I,a} = \sum_{\a} \sum_m 
{\mathcal{S}_I^\a\, (S_{(2n)})_a^m\over \sqrt{\mathcal{S}_0^\a\,
(S_{(2n)})_v^m}}\, \dket{\a,m}~.
\eeq
The fusion algebra for the boundaries $(I,a)$ is the direct product of the
algebras of the irreducible representations of $\Gamma$ (with structure
constants $n_{IJ}^K$) and of the (modified) algebra of the $v,o,s,c$ classes of
SO$(2n)$ described in  Eqs.~(\ref{ven7}-\ref{ven8}). 
\par
Analogously, one constructs the consistent states in a type II theory,
according to the discussion in \secn{sec:fermbsflat}. It is not difficult to
show that the consistent string boundary states constructed according to the
prescription we just described agree with the expressions proposed in
\cite{Roose:2000ay} of the boundary states  of fractional branes for
(non-abelian) orbifolds; the salient coefficients
$\psi^\a_I\sqrt{\sigma(e,g^{(\a)})}$ are already recognised in the bosonic 
expressions \eq{mk13}.
\paragraph{Intersection index}
In~\cite{Douglas:1999hq}, following earlier suggestions 
in~\cite{Berkooz:1996km,Berkooz:1997is}, the  relevance of
the of the Witten index $\mathcal{I}_{IJ}$ for the open strings stretched 
between two branes $I$ and $J$ was pointed out. This index is the one-loop 
amplitude with  boundary
conditions of type $I$ and $J$ in the R$(-)^F$ sector:
\beq
\label{mk20}
\mathcal{I}_{IJ} \equiv Z_{IJ}^{\mathrm{R}(-)^F} =    
{1\over |\Gamma|} \sum_{g\in \Gamma} 
\Tr_{IJ;\mathrm{R}}(\hat g\, (-)^F\,q^{L_0 - {c\over
24}})~.
\eeq   
It is a topological quantity that counts the unpaired chiral massless fermion 
states living on the intersection of the two D-branes. This is the CFT
counterpart of the geometrical intersection form between cycles; more 
precisely,
when the D-branes in question admit a geometrical interpretation as
wrapped higher-dimensional branes, it coincides with the intersection form of
the cycles on which they are wrapped.
\par 
From the cancellation of the non-zero mode contributions to \eq{mk20}, the latter is
$q$-independent, hence topological. This is a generic feature of R$(-)^F$ chiral blocks.
In particular,
\beq
\label{mk21}
\hat\chi^{(X)g^{(\a)}}_e(q)\,\chi_{0,e}^{0,g^{(\a)}}(q) = 
\prod_{l=1}^n (-2\ii\, \sin \pi\nu_{\a,l})~,
\eeq
as follows from \eq{comp6} and \eq{comp9bis} (the fermionic characters in
the R$(-)^F$ sector, $\chi_{0,e}^{0,g^{(\a)}}(q)$, are defined in Appendix
\ref{app:chirblocks}).   Thus we get (see \eq{mk2})
\beq
\label{mk22}
\mathcal{I}_{IJ} =
(-\ii)^n \,\sum_\a (\psi_I^\a)^* \, \psi_J^\a\,
\prod_{l=1}^n (2\sin \pi\nu_{\a,l}) ~.
\eeq
In the tree channel, this amplitude corresponds to ($(-\ii)^n$ times)
the RR contribution (odd spin structure contribution);
it is given by
\beq
\label{mk23}
\mathcal{I}_{IJ} = 
(-\ii)^n\,
\dbra{I;R;\pm}\tilde q^{\frac 1 2 ({L_0 + \tilde L_0 -\frac c {12}})}
\dket{J;R;\pm}~,
\eeq
where we provisionally introduced the states\footnote{%
The tree-channel expression \eq{mk23} can of course also be expressed in terms
of the consistent boundary states of \eq{mk19}, \emph{e.g.},
\beq
\label{mk23a}
\mathcal{I}_{IJ} =
 (\bra{I,s} - \bra{I,c})
\tilde q^{\frac 1 2 ({L_0 + \tilde L_0 -\frac c {12}})}
(\ket{J,v} - \ket{J,o})~,
\eeq
but this way of writing obscures somehow the fact that it is only the 
odd spin structure that is being considered.}
\beq
\label{mk24}
\dket{J;R;\pm} =  
\sum_\b \psi_J^\b\,\sqrt{\sigma(e,g^{(\b)})}\,\dket{\b;R;\pm}~.
\eeq
This gives the same result as in \eq{mk22}, as for the odd spin structure we
have
\beq
\label{mk25}
\dbra{\a;R;\pm}\tilde q^{\frac 1 2 ({L_0 + \tilde L_0 -\frac c {12}})}
\dket{\b;R;\pm} = c_\a\, \delta_{\a\b}~,
\eeq
with $c_\a=1$ if $\a\not=0$ and $c_0=0$, where the index $0$ labels the
conjugacy class of the identity; this is due to the presence of zero-modes in
the untwisted case.
\par
If we were to compute the full string amplitude in the R$(-)^F$  sector, the 
amplitude  $\mathcal{I}_{IJ}$ (like all the amplitudes we have been discussing
in this section) would be multiplied by the  contributions from 
the  directions transverse to the orbifold. 
The full string amplitude thus vanishes
if any Ramond zero-modes are present in these directions.  Nevertheless, when
a geometrical interpretation of the fractional branes is possible,
$\mathcal{I}_{IJ}$ maintains its interest as describing the intersection form
of exceptional cycles.
\subsubsection{Branes on ALE spaces and the McKay correspondence}
Let us next focus attention on complex two-dimensional orbifolds
$\mathbb{C}^2/\Gamma$, with $\Gamma$ a discrete subgroup of SU$(2)$.  Thus,
elements $g\in\Gamma$ act of the complex coordinates $(X^1,X^2)$ in the
defining 2-dimensional representation $\mathcal{Q}$:
\beq
\label{qact}
g:\hskip 0.2cm {X^1\choose X^2} \to \mathcal{Q}(g)\cdot {X^1\choose X^2}~,
\eeq
with $\mathcal{Q}(g)\in\mathrm{SU}(2)$.  Resolving the orbifold singular point
gives rise in this case to   asymptotically locally Euclidean (ALE)
spaces\footnote{%
See for instance \cite{Anselmi:1994sm,Douglas:1996sw,Johnson:1997py} 
for reviews on ALE spaces.}, whose boundary at infinity is $S^3/\Gamma$.
The discrete subgroups of SU$(2)$ are in one-to-one relation to the 
simply-laced ADE Lie algebras. McKay~\cite{mckay} described this
correspondence explicitly as follows (see for instance \cite{Reid99} for a
mathematical review of the classical McKay correspondence and its
generalisations to higher dimensions). 
The tensor product of the defining 2-dimensional representation $\mathcal{Q}$
of $\Gamma$ with an irreducible representation $\mathcal{D}_I$ decomposes as
\beq
\label{mk26}
\mathcal{Q}\otimes \mathcal{D}_I = \oplus_J \widehat A_I^{~J}\, \mathcal{D}_J~.
\eeq
It was observed that
the $r\times r$ matrix $\widehat A$ (where $r$ is the number of irreducible
representations of $\Gamma$) coincides with the adjacency matrix of the 
extended Dynkin diagram corresponding to a simply-laced algebra\footnote{%
In particular, the cyclic $Z_k$ groups are related to the $A_{k-1}$ Dynkin
diagrams, the binary extended dihedral groups $\mathbb{D}_{k}$ to the $D_{k+2}$
diagrams and the  binary extensions of the tetrahedron, octahedron and
icosahedron groups  to, respectively, $E_6$, $E_7$ and $E_8$. }
$\mathcal{G}_\Gamma$ of rank $r-1$.
The extended Cartan matrix of the Lie algebra  is given by 
\beq\label{cext}
\widehat C_I^{~J} =
2\delta_I^{~J} - \widehat A_I^{~J}~.
\eeq
\par
Take an element $g^{(\a)}$ in the $\mathcal{Q}$ rep\-res\-en\-ta\-ti\-on,
and diagonalise it:
\[
\mathrm{diag}(\ee^{2\pi\ii\nu_\a},\ee^{-2\pi\ii\nu_\a})\ .
\]
The character of
$\mathcal{Q}$ evaluated on $g^{(\a)}$ is thus $2\cos 2\pi\nu_\a = 2 - 4\sin^2
\pi\nu_\a$. Taking traces and using the orthogonality of characters,
one then derives from McKay's relation \eq{mk26} that
\beq
\label{mk27}
\widehat C_I^{~J} = 
\sum_\a \psi_I^\a\,(\psi_J^\a)^*\, 4\sin^2 \pi\nu_\a~.
\eeq
The quantities $\psi_I^\a$ introduced in \eq{mk13a} are the eigenvectors
of the extended Cartan matrix $\widehat C_I^{~J}$, and eigenvalues of the latter are in
correspondence with conjugacy classes of $\Gamma$; the eigenvalues are
$\sigma_\a = 4\sin^2\pi\nu_\a$. 
\par
As is well known from the work of Kronheimer \cite{kronheimer}, the Dynkin
diagram of the Lie algebra $\mathcal{G}_\Gamma$  plays a fundamental
geometrical role. Resolving the singularity of the $\mathbb{C}^2/\Gamma$ space
gives rise to a smooth ALE space $\mathcal{M}_\Gamma$,  which has a non-trivial
middle homology group $H_2(\mathcal{M}_\Gamma,\mathbb{Z})$ of dimension $r-1$,
the rank of $\mathcal{G}_\Gamma$. The (symmetric)
intersection matrix of the $r-1$ exceptional two-cycles $c_{i}$ ($i=1,\ldots
r-1$), is given by the non-extended
Cartan matrix $C_{i}^{~j}$  of $\mathcal{G}_\Gamma$. Thus the 2-cycles 
$c_{i}\in H_2(\mathcal{M}_\Gamma,\mathbb{Z})$ correspond to the simple roots
$\a_{i}$ of $ \mathcal{G}_\Gamma$. Furthermore,  the cycle $c_0 = -\sum_{i}
d_{i} c_{i}$ is associated to  the negative of the highest root,  $\a_0 =
-\sum_{i} d_{i}\a_{i}$, \emph{i.e.}, to the extra node in the \emph{extended} Dynkin
diagram. Here $d_{i}$ are the Coxeter numbers and $d_0=1$; in the McKay
relation the $d_I$'s correspond to the dimensions of the irreducible
representations of $\Gamma$.  The intersection matrix of the cycles
$\{c_I\}=\{c_0,c_{i}\}$ is then  the negative of the \emph{extended} Cartan
matrix $\widehat C_I^{~J}$.  
\paragraph{Fractional branes as wrapped branes}
The fractional D$p$-branes of type  $\ket{I}$ correspond to D($p+2$)-branes
wrapped on an  \emph{integer} basis of cycles  $c_{I}\in
H_2(\mathcal{M}_\Gamma,\mathbb{Z})$, in the collapsing limit in which
these cycles shrink\footnote{%
The supergravity classical solution corresponding to a fractional 
D-brane can be retrieved~\cite{Bertolini:2000dk} in accordance with this
interpretation.}.   
\par
The first test  of this interpretation is provided by the the topological
partition function~\cite{Takayanagi:2000rd,Roose:2000ay}  $\mathcal{I}_{IJ} =
Z_{IJ}^{\mathrm{R}(-)^F}$  between D$p$-branes of types $I$ and $J$, that
computes the intersection of the cycles $c_I$ and $c_J$.  In the case of
$\mathbb{C}^2/\Gamma$, \eq{mk23} gives
\beq
\label{ mk28}
\mathcal{I}_{IJ} = 
-\sum_\a \psi_I^\a\,(\psi_J^\a)^*\, 4\sin^2 \pi\nu_\a
= - \widehat C_{IJ} = c_I\cdot c_J~,
\eeq 
in full correspondence with the above discussion of the non-trivial 2-cycles
of $\mathcal{M}_\Gamma$. 
\par
The geometrical interpretation of the fractional branes as wrapped branes must
furthermore be consistent with their masses and charges  (both twisted and
untwisted).
\par
The boundary states for the consistent fractional branes encode  the corresponding D-brane
couplings to the RR ($p+1$)-form fields $A^{(\a)}_{p+1}$ that arise in the
untwisted ($\a=0$) and twisted ($\a\not= 0$) sectors of type II string theory
on  $\mathbb{R}^{1,5}\times \mathbb{C}^2/\Gamma$.  The RR components of the
boundary states in the orbifold CFT read explicitly
\beq
\label{mk29}
\ket{I;R:\pm} = \psi_I^0 \dket{0;R;\pm} + \sum_{\a\not= 0} \psi_I^\a\, 
2\sin \pi\nu_\a \dket{\a;R;\pm}~.
\eeq
Let us focus first on the branes of type $\ket i$, \emph{i.e.}, those
corresponding to non-trivial representations of $\Gamma$. The charges of such
branes  with respect to the twisted fields $A^{(\a)}_{p+1}$, with $\a\not=0$, are
encoded in the matrix
\beq
\label{twistcharge1}
Q_i^{~\a} = \mu^{(\a)}_p\,\psi_{i}^{~\a}\,\sqrt{\sigma^\a}~,\hskip 0.3cm
(i,\a=1,\ldots r-1)~,
\eeq         
where $\sigma^\a = 4\sin^2 \pi\nu_\a$ are the non-zero eigenvalues of the
\emph{extended} Cartan matrix; we have included the overall normalisation 
$\mu^{(\a)}_p$ of \eqn{twisttmu}. 
\par
Considering the theory on the resolved ALE space, the geometrical counterparts
of the twisted gauge fields~\cite{Douglas:1997xg} are the Kaluza-Klein
$p+1$-forms  $\mathcal{A}^{(i)}_{p+1}$ that come from the harmonic
decomposition 
\beq
\label{twistcharge2}
A_{p+3}=\sum_{i} \mathcal{A}_{p+1}^{(i)}\wedge \omega_{i}~.
\eeq 
Here, $A_{p+3}$ is the ten-dimensional $A_{p+3}$ RR form and the $\omega_i$'s
are the normalisable anti-self-dual $(1,1)$-forms, Poincar\'e dual
to the  cycles $c_{i}$, 
so that
\beq
\label{twistcharge3}
\int_{\mathcal{M}_\Gamma} \omega_i\wedge\omega_j=\int_{c_{i}}\omega_j =
 c_i\cdot c_j= -C_{ij}~. 
\eeq
Notice that both the $\mathcal{A}_{p+1}^{(i)}$ and the $A^{(\a)}_{p+1}$ fields
depend only on the six coordinates transverse to the orbifold, the
first ones because of their KK origin, the second ones because of the absence
of momentum along the orbifold directions in the twisted sectors.
\par
A D$(p+2)$-brane wrapped on the cycle $c_i$, which corresponds to the 
fractional brane $\ket i$ in the orbifold limit, is charged under the fields 
$\mathcal{A}^{(j)}_{p+1}$ because of its Wess-Zumino coupling
\beq
\label{twistcharge4}
\mu_{p+2}\, \int_{\mathrm{D}(p+2)}A_{p+3} = 
\mu_{p+2}\, \int_{\mathrm{D}p}\mathcal{A}^{(j)}_{p+1}\,
\int_{c_i}\omega_j =
-C_{ij}\; \mu^{(\a)}_p\, \int_{\mathrm{D}p}\mathcal{A}^{(j)}_{p+1}~.
\eeq
Notice that the usual D$(p+2)$ charge $\mu_{p+2}$, which is appropriate for the
branes in the smooth resolved space $\mathbb{R}^{1,5}\times\mathcal{M}_\Gamma$,
equals the twisted charge normalisation $\mu^{(\a)}_p$, as follows from 
Eqs.~(\ref{RRcharge},\ref{twisttmu}).   Comparing the charge matrices $Q_{i\a}$ of
\eq{twistcharge1} to $\mu^{(\a)}_p \,C_{ij}$ encoding the charges with respect to 
the ``geometric'' fields $\mathcal{A}^{(j)}$,
one is led to conclude that the two bases of
($p+1$)-forms are related by
\beq
\label{twistcharge5}
\mathcal{A}^{(i)}_{p+1} = -\sum_{j,\a} 
(C^{-1})^{ij} \psi_j^{~\a}\sqrt{\sigma^\a} A^{(\a)}_{p+1}~.
\eeq
In a matrix notation we write this change of basis as
$\mathcal{A}_{p+1} = C^{-1}  \psi^T \sqrt{\Sigma} A_{p+1}$.  This relation is
consistent with the bulk kinetic terms for these fields. Indeed,  from a
canonical kinetic term for $A_{p+3}$,  under the decomposition
\eq{twistcharge4} and  making use of \eq{twistcharge3}, we get  a
six-dimensional bulk kinetic term
\beq
\label{twistcharge6}
{1\over 2} \int_{\mathbb{R}^{1,5}} \bar \mathcal{F}^{(i)}_{p+2}\, 
\wedge C_{ij}\,{}^*\mathcal{F}^{(j)}_{p+2}
=
{1\over 2} \int_{\mathbb{R}^{1,5}} \bar F^{(\a)}_{p+2}\wedge  
{}^* F^{(\a)}_{p+2}~,
\eeq
where in the second step we changed basis according to \eq{twistcharge5}. The
fact that the kinetic terms become canonical in terms of the
twisted fields $A^{(\a)}_{p+1}$ is an non-trivial consistency  requirement:
closed string twisted sectors are mutually orthogonal and as such cannot produce a
non-diagonal effective kinetic term. The fact that the change of basis 
\eq{twistcharge5} indeed performs this diagonalisation is not
obvious a priori. A proof involves the fact that (with $\a,\b\not=0$)
\beq
\label{twistcharge7}
\sum_{i,j} 
(\psi^*)_\a^{~i}\, (C^{-1})_i^{~j} \psi_j^{~\b} = {1\over\sigma^\a}\,
\delta^{\a}_{~\b}~.
\eeq 
This property can be proven using that the $\psi^\a_I$ are
eigenvectors of the \emph{extended} Cartan matrix\footnote{
Contrary to one's naive first guess,
the closed string basis $A^{(\a)}_{p+1}$ is not related to the
``geometrical'' one $\mathcal{A}^{(i)}_{p+1}$ by the change of basis
$ \mathcal{A}_{p+1} =q^T\, \Lambda^{-1/2} A_{p+1}$,  where
$q$ is the (orthonormal) matrix of eigenvectors of the Cartan matrix  and
$\Lambda$  the diagonal matrix containing its eigenvalues.
Although this transformation makes the kinetic terms canonical,
the true closed-string basis requires a further
transformation $U=\Lambda^{-1/2}\,q\,\psi^TW\Sigma^{1/2}$. Since the latter is unitary the
canonical form obtained by the first transformation is preserved.
} with corresponding
orthogonality properties 
\beq
\label{twistcharge8}
\sum_{I=0}^{r-1} (\psi^\a_{~I})^* \psi_I^{~\b} = \delta^{\a\b}~, \hskip 1cm
\sum_{\a=0}^{r-1} (\psi_I^{~\a})^* \psi^{\b}_{~J} = \delta_{IJ}~.
\eeq
In fact, the latter follow 
from the properties of the character matrix $\rho$ (see Appendix
\ref{app:discrete}), in view of $\psi^\a_I = \sqrt{n^\a/|\Gamma|}\,\rho^\a_I$.
\par  
The twisted charges of the fractional brane $\ket 0$ associated to the trivial
representation are described by the vector  $Q^\a = \mu^{(\a)}_p\,\psi_0^{~\a}
\sqrt{\sigma^\a}$ $ =-\mu^{(\a)}_p\,\sqrt{\sigma^\a/|\Gamma|}$.  They are
correctly accounted for by assuming that this state corresponds to a brane
wrapped on the cycle $c_0 = -\sum d_i c_i$, as is suggested by McKay's
correspondence  and easily shown using again the orthogonality properties
\eq{twistcharge8}.  
\par  
The fractional branes $\ket I$ are furthermore charged with respect  to the RR
untwisted  ($p+1$)-form $A_{p+1}$. It follows from \eq{mk29} and  from the
discussion around \eq{tmugamma} that their untwisted charges are 
\beq
\label{twistcharge9} 
Q_I = \mu^{(\Gamma)}_{p}\, \frac{d_I}{\sqrt{|\Gamma|}} = 
\mu_p \, \frac{d_I}{|\Gamma|}~.
\eeq
From the geometric point of view, the untwisted charges $Q_I$ arise
from the Wess-Zumino coupling
\beq
\label{twistcharge10} 
\mu_{p+2}\, \int_{\mathrm{D}(p+2)} 
A_{p+1}\wedge (2\pi\a' \mathcal{F} + B)
= \mu_{p+2}\,\int_{\mathrm{D}p} A_{p+1} \int_{c_I} (2\pi\a' \mathcal{F}+ B)~, 
\eeq
since the orbifold  conformal field theory is obtained with a non-trivial 
$B$-background 
\cite{Aspinwall:1995zi,Douglas:1997xg,Nahm:1999ps,Wendland} 
localised on the vanishing cycles.  Consider indeed the branes of type $\ket
i$. The charges  $Q_i$ of \eqn{twistcharge9} are retrieved from 
\eqn{twistcharge10} by wrapping a brane with no world-volume gauge field
$\mathcal{F}$ and assuming 
\beq
\label{twistcharge11}
B = -\frac{\mu_p}{\mu_{p+2}}{1\over |\Gamma|}
    \sum_i d_i (C^{-1})^{ij} \omega_j 
  = \frac{\mu_p}{\mu_{p+2}}{1\over |\Gamma|}
    \sum_i d_i \omega^i~.
\eeq 
The (1,1)-forms $\omega^i=-(C^{-1})^{ij} \omega_j$ satisfy the property that
$\int_{c_i}\omega^j = \delta^i_j$. Notice also that  $\mu_p/\mu_{p+2} =
4\pi^2\a'$.  The expression of $B$ given above agrees with the one obtained  
in \cite{Wendland} (sec. 4) as a local limit from the requirement of  consistent
embeddings of orbifold CFT's into the part of moduli space of
$\mathcal{N}=(4,4)$ theories describing K3 sigma-models.  
\par
Consider however the brane of type $\ket 0$.  To get from the Wess-Zumino term
\eqn{twistcharge10} the untwisted charge $Q_0 = \mu_{p}/|\Gamma|$,  in addition
to having the $B$-background  \eqn{twistcharge11} we must be wrapping a
D$(p+2)$ brane with a non-trivial open string $\mathcal{F}$ field 
\cite{Douglas:1997xg,Bergman:1999kq,Dasgupta:1999wx}.  Indeed, the
$B$-contribution from wrapping over the cycle $c_0$ would be given by $-
\mu_p\,\sum_i d_i^2/|\Gamma|$  $= \mu_p (1-|\Gamma|)/|\Gamma|$. To cancel the
extra negative charge we need a  $\mathcal{F}$ background localised on $c_0$.
Such a  background is quantised, as its Chern class is
$\int_{c_0}\mathcal{F}=2\pi k$,  with $k\in\mathbb{Z}$. In our case, we just
have to take $k=1$  to get from  \eqn{twistcharge10} the sought for
contribution of   $4\pi^2\a'\mu_{p+2}= \mu_{p}$. 
\par
From the above discussion it follows that a superposition of fractional branes
 that transforms in the regular representation, namely  $\sum_I d_I \ket
I$ according to the decomposition \eq{mm1}, is not charged under the twisted
sector gauge fields and has
untwisted charge $\mu_p \sum_I(d_I)^2/|\Gamma|=\mu_p$. Thus it is
degenerate with a bulk D$p$-brane located at the  orbifold fixed point,
as expected. 
\par 
\paragraph{Relation to branes on ALE spaces at their Gepner points} 
In the present subsection we have been studying D-branes on a
$\mathbb{C}^2/\Gamma$ orbifold. Geometrically, such orbifolds are degenerate
ALE spaces, the two-cycles $c_{i}$ having zero size. However, the orbifold CFT
is non-singular due to a non-zero $B$-flux through these vanishing cycles
\cite{Aspinwall:1995zi,Nahm:1999ps}.   
In the recent papers \cite{Lerche:2000uy,Lerche:2000jb},
D-branes have been studied at a different point of the ALE moduli space,
the Gepner point. The CFT in question may be formulated as a coset model based on
\beq
\label{gep3}
\left(\frac{\mathrm{SU}(2)_{h-2}}{\mathrm{U}(1)}\times
\frac{\mathrm{SL}(2)_{h+2}}{\mathrm{U}(1)}\right){\Big /} \mathbb{Z}_h~,
\eeq
where $h$ is the Coxeter number of ${\cal G}_\Gamma$.
By constructing boundary states and computing their intersection index, the
interpretation in terms of branes wrapping two-cycles was derived. 
The fact that the intersection index of \cite{Lerche:2000uy,Lerche:2000jb}
agrees with the one of the orbifold theory described here is no surprise: given
the fact that string theory on an ALE space preserves sixteen supersymmetries,
one expects the spectrum of BPS D-branes to be the same in both theories.
\sect{Discrete torsion}
\label{sec:discretetorsion}
In \cite{Vafa:1986wx}, it was noted that it is consistent with modular
invariance to include certain phases in the orbifold partition function
$Z(\tau,\bar\tau)$ on a torus with modular parameter $\tau$:
\beq\label{partfionDT}
Z=\frac1{|G|}\sum_{\begin{array}{c}{g,h\in
G}\\{[g,h]=e}\end{array}}Z(g,h)\e(g|h)~.
\eeq 
The phases $\e(g|h)$ are called discrete torsion.
\par
Higher loop modular invariance implies the following consistency conditions on
$\e(g|h)$ \cite{Vafa:1986wx}:
\bea
\e(gh|k)&=&\e(g|k)\e(h|k)~;\label{DTghk}\\
\e(g|h)&=&\e(h|g)^{-1}~;\label{DTgh}\\
\e(g|g)&=&1~.\label{DTgg}
\eea
\par
From the Hamiltonian (rather than the functional integral) point of view, 
$\e(g|h)$ changes the phase of the operator corresponding to $g$ in the sector 
twisted by $h$. States in ${\cal H}_h$ that were invariant under the original
action of $N_h$, will only survive the orbifold projection with discrete
torsion if $\e(g|h)=1$ for all $g\in N_h$.
\par
D-branes in theories with discrete torsion were introduced in
\cite{Douglas:1998xa}, at least for abelian orbifold groups. We shall review
this construction,  generalised to non-abelian $G$,%
\footnote{Recently, this generalization has also been considered in
\cite{Aspinwall}.} and construct boundary states for D-branes localised
at an orbifold fixed  point.  
\par
To study D-branes in an orbifold theory, one starts with a $G$-symmetric
configuration of D-branes on the covering space of the orbifold. In addition to
the space-time action $r(g)$, one considers a representation $\gamma(g)$ of $G$
in the gauge group of the D-branes. The fields $\phi$ living on the D-branes
are then projected as 
\beq
\g^{-1}(g)\phi\g(g)=r(g)\phi~.
\eeq
In theories without discrete torsion, $\g$ is a genuine representation of the
orbifold group. In theories with discrete torsion, $\g$ is taken to be a
projective representation \cite{Douglas:1998xa}:
\beq\label{projrep}
\g(g)\g(h)=\omega(g|h)\g(gh)~.
\eeq
As we will discuss soon, the {\it factor system}\, $\omega(g|h)$, which
consists of non-zero complex numbers, is related to the discrete torsion
phases  $\e(g|h)$. Before we come to that, we briefly study  factor systems by
themselves. 
\par
From
\beq
\g(ghk)=\omega(gh|k)^{-1}\g(gh)\g(k)=\omega(gh|k)^{-1}\omega(g|h)^{-1}\g(g)
\g(h)\g(k)
\eeq
and
\beq
\g(ghk)=\omega(g|hk)^{-1}\g(g)\g(hk)=\omega(g|hk)^{-1}\omega(h|k)^{-1}\g(g)
\g(h)\g(k)~,
\eeq
we deduce the {\it associativity condition}
\beq\label{associativity}
\omega(gh|k)\omega(g|h)=\omega(g|hk)\omega(h|k)~.
\eeq
Rescaling the representation matrices by $\g(g)=\b(g)\g'(g)$, we find the
{\it equivalent factor system}
\beq\label{equivfactor}
\omega'(g|h)=\frac{\b(gh)}{\b(g)\b(h)}\omega(g|h)~.
\eeq
More abstractly, the associativity condition \eq{associativity} and the
equivalence relation \eq{equivfactor} are combined in the statement that the
factor system is a two-cocycle in the group cohomology $H^2(G,{\rm U(1)})$. 
\par
Consider two commuting elements $g,h\in G$. From the factor system $\omega$ we
define 
\beq\label{epsomega}
\e(g|h)=\omega(g|h)\omega(h|g)^{-1}~;\ \ \ [g,h]=e~.
\eeq
As the notation suggests, $\e(g|h)$ will be identified with a discrete torsion
phase. We now check that this identification could make sense. Doing these
checks, we shall extensively use the fact that $[g,h]=e$. Note, however, that
$\e(g|h)$ is only defined for commuting group elements, so that our discussion
is in no way restricted to abelian orbifold groups. First, taking determinants
in \eq{projrep} and the same equation with $g$ and $h$ reversed, we find that
$|\e(g|h)|=1$. Second, it is easy to check that \eq{epsomega} is invariant
under the equivalence relation \eq{equivfactor}. Third, \eq{DTgh} and \eq{DTgg}
are manifestly satisfied. Finally, \eq{DTghk} follows from  \eq{associativity}.
\par
Of course, these consistency conditions alone do not imply that \eq{epsomega}
makes sense. For a certain abelian orbifold, \eq{epsomega} was derived in
\cite{Douglas:1998xa} by considering a specific genus one diagram with one
boundary. In what follows, we shall check the relation \eq{epsomega} by
factorising a cylinder amplitude between two D-branes in the closed string
channel. We take the D-branes to be localised at a fixed point of the
orbifold.%
\footnote{For D-branes extended in the orbifold directions, the
relation between discrete torsion and projective representations is more subtle
\cite{Gaberdiel:2000fe}.}  We shall show that from  ${\cal H}_h$ only states
invariant under the $N_h$ projection {\it with  $\e(g|h)$ included} contribute
to the amplitude. The factorization in the closed string channel will be
achieved by constructing boundary states for the D-branes with discrete
torsion. It will then be sufficient to check that these boundary states are
well-projected.
\par
The boundary states for fractional branes in the theory without discrete
torsion were given in \eq{mk19}. The Ishibashi components are weighted by
factors proportional to the character $\rho_I^\alpha$, where $\a$ denotes the
twisted sector to which the Ishibashi component belongs and $I$ is the
representation of the orbifold group associated to the fractional brane. The
consistent boundary states in a theory with discrete torsion are obtained by
replacing $\rho_I^\alpha$ by a projective character $\r(g^{(\a)})$ of a group
element in the conjugacy class ${\cal C}^\a$.%
\footnote{As we are about to discuss, projective characters are not class
functions, so the freedom in choosing $g^{(\a)}\in{\cal C}^\a$ leads to an
ambiguity. Since the matrices acting on Chan-Paton factors have determinants of
modulus one, the ambiguity is just a phase factor, which is irrelevant for our
purposes.} This raises the following question. The Ishibashi states
\eq{invIshi} were invariant under the orbifold action {\it without} discrete
torsion. When discrete torsion is turned on, some of these Ishibashi states
will no longer be physical. How do the boundary states we have just constructed
know about that? To answer this question, we have to study projective
characters in a bit more detail. 
\par
The character $\r$ associated to a projective representation $\g$ is defined as
\beq\label{charproj}
\r(h)=\tr\g(h)~.
\eeq  
Unlike the character of genuine representations, the character of a projective
representation is not a class function. In general, we have
\beq\label{characterconj}
\r(ghg^{-1})=\omega(gh|g^{-1})^{-1}\omega(g|h)^{-1}\omega(g^{-1}|g) \omega(h|e)
\r(h)~.
\eeq
If $g$ and $h$ commute, the left hand side of \eq{characterconj} reduces to
$\r(h)$, so $\r(h)$ can only be non-zero if the product of $\omega$'s on the
right hand side of \eq{characterconj} equals one. Because $g$ and $h$ commute,
we can use associativity to write
\beq
\omega(gh|g^{-1})^{-1}=\omega(h|g)\omega(h,e)^{-1}\omega(g,g^{-1})^{-1}~,
\eeq
such that the product of $\omega$'s on the
right hand side of \eq{characterconj} becomes just
\beq
\omega(gh|g^{-1})^{-1}\omega(g|h)^{-1}\omega(g^{-1}|g)
\omega(h|e)=\e(h|g)\e(g^{-1}|g)=\e(h|g)~.
\eeq
Here, $\e$ is as defined in \eq{epsomega} and the fact that $ \e(g^{-1}|g)=1$
follows easily from \eq{DTghk} and \eq{DTgg}. Thus, we can conclude that if
a $g\in N_h$ exists such that $\e(g|h)\neq1$, then
\beq
\r(h)=0~.
\eeq 
For our boundary states with discrete torsion,  this property implies that
Ishibashi states that  are not physical in the theory with discrete torsion  do
not contribute to the consistent boundary states.  This is a direct consistency
check for the relation between discrete  torsion and projective representations
for D-branes at an orbifold fixed point. 
\section{Conclusions}
In this paper we have shown how to generalise Cardy's construction of consistent boundary
states in the case of complex orbifolds. The distinctive feature is that the operation 
taking Ishibashi states (cohomology) into Cardy states (homology) involves a generalised
discrete Fourier transform. The latter naturally implements the known identification of
fractional D-branes at the fixed point with irreducible representations of the orbifold 
group. In particular, we put quite some effort in the proper identification of the twisted RR 
form basis that is natural from the geometric point of view, {\em i.e.}, the 
basis obtained after KK
reduction on the Poincar\' e duals of the vanishing cycles. As such, the constructed boundary
states were shown explicitly to allow an interpretation as branes wrapping cycles with some
particular B-flux. Although attention was mainly focused on the case of complex orbifold
surfaces, it is obvious that the Cardy construction also applies in more involved cases. 

In the final section we moved attention to orbifolds with discrete torsion. The boundary states
describing pointlike branes on these orbifolds paved the way to establish the link between
D-branes as projective representations on the one hand and discrete torsion phases in the
closed string partition function on the other hand.
\acknowledgments
We are grateful to D.~Berenstein, P.~De~Smet, M.~Frau, M.~Gaberdiel, L.~Gallot,
J.~Harvey, D.~Kutasov, A.~Lerda,  E.~Martinec, I.~Pesando, R.~Russo and S.~Sethi
for interesting discussions.  We would like to thank the University of Torino
(B.C.) and the organisers of the "Semestre Cordes" at the Centre Emile Borel
of the Institut Henri Poincar\' e (F.R.) for hospitality during the initial and
final stages of this work, respectively.  This work was supported by DOE grant
DE-FG02-90ER-40560, NSF grant  PHY-9901194, by the European Commission TMR
programme ERBFMRX-CT96-0045 and by the RTN programme HPRN-CT-2000-00131. F.R. was
supported in part by a Marie Curie Fellowship of the European Community
Programme "Improving human research potential and the socio-economic knowledge base" under
contract number HPMT-CT-2000-00163.  During part of this work, B.C.
was Aspirant FWO-Flanders.
\appendix
\section{Boson boundary state}
\label{bosbsflat}
This Appendix reviews the construction of consistent boundary states
in the CFT of free bosons.~\cite{Recknagel:1998sb,Fuchs:1998fu}  

Consider first a single boson on a circle of radius $R$.  The sub-algebra
${\cal A}$ is then generated by $\partial X$, and contains the identity
operator.  In the holomorphic and anti-holomorphic sectors, the Fourier modes 
$\alpha_n$ and $\tilde \alpha_n$ are defined by
\beq
\partial X(z) = -\ii\sqrt{\frac{\a'}{2}}\sum_{m=-\infty}^\infty
\a_mz^{-m-1}~;\hskip 0.2cm
\bar\partial X(\bar z)= -\ii\sqrt{\frac{\a'}{2}}\sum_{m=-\infty}^\infty
\tilde\a_m\bar z^{-m-1}\label{dXinmodes}
\eeq
and obey the algebra
$[\a_m,\a_n]=[\tilde\a_m,\tilde\a_n]=m\,\d_{m+n}$
On a generic circle this CFT has an infinite number of highest weight states
$\ket{(k,w)}$ ($k,w \in \mathbb{Z}$). These have no oscillators excited and are
thus defined by
\begin{eqnarray}
\hat p \ket{(k,w)} &=& \frac k R \ket{(k,w)} \ ; \\
\hat w \ket{(k,w)} &=& w \ket{(k,w)} \ ; 
\end{eqnarray}
with $\hat p = (\alpha_0 + \tilde\alpha_0)/\sqrt{2\a'}$, and $\hat w =
\sqrt{\frac{\a'}2}(\alpha_0 - \tilde\alpha_0)/R$.  
\par
Taking an automorphism $\Omega$ of ${\cal A}_R$, one may impose the following
gluing conditions at world-sheet boundaries (see \eq{BSbc}):
\begin{equation}\label{genbndcnd}
\left(\alpha_n + \Omega( \tilde \alpha_{-n})\right)\dket{i}_\Omega = 0\ .
\end{equation}
On $\dket{i}_\Omega$ the left-moving and right-moving closed string Virasoro
operators get identified: $(L_0^c - \tilde L_0^c)\dket{i}_\Omega = 0$.

The special case in which $\Omega$ equals the identity corresponds to Neumann
boundary conditions, while Dirichlet boundary conditions are realised by
$\Omega(\bar \partial X) =  - \bar\partial X$.  The generalised Ishibashi
states solving \eq{genbndcnd} 
for the special cases of Dirichlet or Neumann gluing conditions will be
denoted by $\dket{i}_D,\dket{i}_N$, respectively.
\par
With Neumann boundary conditions, the Ishibashi states are
\begin{equation}\label{Nishist}
\dket{(0,w)}_N =
\exp\left(-\sum_{n = 1}^{\infty}\frac 1 n \alpha_{-n}\tilde\alpha_{-n}
\right)\ \ket{(0,w)}
\end{equation}
with corresponding Ishibashi characters \eq{ishichar}
\begin{equation}
\chi_{N,w}(\tilde q) = \frac{\tilde q^\frac{R^2 w^2}{4\alpha
'}}{\eta(\tilde q)}\ .
\end{equation}
The Ishibashi states of \eq{Nishist} satisfy the orthogonality condition
\eq{ortho}. 
Notice that the highest weight representations $j$ of the chiral preseved 
chiral algebra are labeled by the winding
number $j=(0,w)$ here. The orthonormal set of basis vectors $\ket{j;N}$ becomes
$\ket{(0,w);\{m_n\}}$ in this concrete setting, where $m_n$ denote the
$\alpha_{-n}$ oscillator numbers.
\par
As to the boundary state/D-brane correspondence, one has to specify which open
strings/boundary conditions are 
physically sensible. With Neumann  boundary conditions along the compact $X$ 
direction, one allows open strings to carry non-zero momentum along that 
direction, so that the corresponding open 
string character $\chi(q)_N={\rm Tr}_N(q^{L_0^{o}})$ contains a momentum sum. 
Moreover, 
one can turn on a Wilson line $A_X = \frac \theta {2\pi R}$
with $\theta \in [0,2\pi[ $; this is equivalent to shifting the 
open string momenta:
$ \frac n R \rightarrow \frac n R - \frac \theta {2\pi R}$.
Thus the generic open string character $\chi(q)_{N,\theta}$ reads

\begin{eqnarray}
\chi(q)_{N,\theta} & = & 
\sum_{n\in \mathbb{Z}}\frac{ q^{(\frac n R - \frac \theta {2\pi R})^2}}
{\eta(q)}\nonumber\\
& = & \frac R {\sqrt{2\alpha '} } \sum_{w\in \mathbb{Z}} 
\ee^{\ii\theta w}\frac{\tilde q^\frac{R^2 w^2}{4\alpha'}}{\eta(\tilde q)}
= \frac R {\sqrt{2\alpha '} } \sum_{w\in \mathbb{Z}} \ee^{\ii\theta w}\
\chi_{N,w}(\tilde q)~,
\end{eqnarray}
where in the second line we performed a Poisson resummation.
The corresponding consistent boundary states are
\begin{equation}\label{thetaNbs}
\ket{\bf \theta}_N = \left(\frac {R} {\sqrt{2\alpha '}}\right)^{1/2} \sum_{w
\in \mathbb{Z}} e^{\ii\theta w} \dket{(0,w)}_N\ .
\end{equation}
Notice that these states take the form of \eq{coeffS}, with 
$S_\theta^w =\frac R {\sqrt{2\alpha '} }\,e^{\ii\theta w}$.
The role of the distinguished Cardy's state $\ket{\bf 0}$ is clearly played by
the state corresponding to no Wilson line, $\ket{\bf 0}_{N}$.

\par
Turning to Dirichlet boundary conditions, we take $\Omega(\bar \partial X) = -
\bar \partial X$. This is evidently an implementation of T-duality (which is a
one-sided parity transformation). Instead of \eq{Nishist} one now has
generalised Ishibashi states 
\begin{equation}\label{Dishist}
\dket{(k,0)}_D =
\exp\left(\sum_{n = 1}^{\infty}\frac 1 n \alpha_{-n}\tilde\alpha_{-n}
\right)\ \ket{(k,0)}
\end{equation}
solving \eq{genbndcnd} with corresponding Ishibashi characters \eq{ishichar} 
\begin{equation}
\chi_{D,k} = \frac{\tilde q^\frac{\alpha ' k^2}{4 R^2}}{\eta(\tilde q)}\ .
\end{equation}
\par
D-branes localised along the $X$ direction make sense, and the corresponding
boundary states are constructed as follows. Since the Dirichlet boundary
condition at both ends sets the open string momentum along $X$ to zero but
allows for non-zero winding, which must accordingly be summed over in the  open
string character. For an open string stretched between two D-branes at a distance
$\Delta x$ from each other, the character ${\rm Tr}_D(q^{L_0^{o}})$ is 
\begin{eqnarray}
\chi(q)_{D,\Delta x} & = & 
\frac{ q^\frac{(2\pi R w + \Delta x)^2}{4\pi^2\alpha '}}
{\eta(q)}
\nonumber\\
& = & \frac{\sqrt{\alpha '}}{\sqrt 2 R } \sum_{k\in \mathbb{Z}} 
\ee^{\ii \Delta x \frac k R}
\frac{\tilde q^\frac{\alpha ' k^2}{4 R^2}}{\eta(\tilde q)}
= \frac{\sqrt{\alpha '}}{\sqrt 2 R }  \sum_{k\in \mathbb{Z}} 
\ee^{\ii \Delta x \frac k R}\chi_{D,k}(\tilde q)~,
\end{eqnarray}
where in the second line we went to the tree channed by Poisson resummation.
The corresponding consistent boundary states describe D-branes localised at
fixed positions $x$ along the circle:
\begin{equation}\label{xDbs}
\ket{x}_{D} = \left(\frac{\sqrt{\alpha '}}{\sqrt 2 R }\right)^{1/2}
\sum_{k \in \mathbb{Z}} \ee^{\ii x \frac k R}\dket{(k,0)}_D\ .
\end{equation}
Also in this case, these state assumes the form of \eq{coeffS}.
The distinguished state $\ket{\bf 0}$ is of course simply 
$\ket{x=0}_{D}$.
The close analogy
between \eq{thetaNbs} and \eq{xDbs} reflects the fact that turning on Wilson
lines results in shifting positions of branes in the T-dual picture.
\par
The above discussion was restricted to the case of one free boson so as to
simplify the discussion. As a nice result, one verifies that the ratio of the
coefficients in front of $\ket{\theta=0}_N$ and $\ket{x=0}_D$ is
$(2\pi R)/(2 \pi \sqrt{\alpha '}$
thus confirming that the relative tensions differ by
$\frac{1}{2\pi\sqrt{\alpha '}}$~\cite{Polchinski:1995mt}.
\par
Moreover, regarding the states in \eq{thetaNbs}  and \eq{xDbs} as constituents,
the generalisation to the case of several bosons is obvious and
consists of taking tensor products of the appropriate constituents.
\par
To study the decompactification limit $R\rightarrow\infty$, note that the
states $\ket{(k,0)}$ appearing in \eq{Dishist} are normalised to one. Thus
they differ from the states $e^{\ii k\hat x}\ket{0}$ that are usually
considered in string theory by a factor of $1/\sqrt{2\pi R}$. Taking into
account the $1/\sqrt R$ factor manifest in \eq{xDbs}, the $R$-dependence will
be just right to turn the discrete momentum sum in \eq{xDbs} into a momentum
integral. This momentum integral then gives the usual delta function
localising the brane in the transverse directions (see, for instance,
\cite{Billo:1997eg,DiVecchia:1997pr}).
\section{Chiral blocks}
\label{app:chirblocks}
In this Appendix we collect the explicit expressions for the chiral blocks of
orbifolds of free bosonic and fermionic theories. As discussed at the beginning
of \secn{sec:chiral}, we consider orbifolds of $\mathbb{C}^n$ by a point group
$\Gamma$, and in each sector $(g,h)$ we can diagonalise simultaneously both $h$
and $g$, as in \eq{comp2} and \eq{comp3}.  As a result, one can consider
separately each complex direction.  We can therefore limit ourselves to
describe here the chiral blocks 
\beq
\label{appc1}
\chi_h^g(q)=\Tr_{\mathcal{H}_h}\, g\, q^{L_0 - {c\over 24}}
\eeq
for a single complex bosonic field $X$ and a fermionic one $\psi$, for which a
sector denoted by $(g,h)$ is characterised by a twist by $h$ which amounts to
\beq
\label{appc2}
X\in\mathcal{H}_h:\hskip 0.2cm
X(\tau,\sigma+2\pi) = \ee^{2\pi\ii\nu}X(\tau,\sigma)
\eeq
and by an action of $g$ given by
\beq
\label{appc3}
g:\hskip 0.2cm X\mapsto \ee^{2\pi\ii\nu'} X
\eeq
(and similarly for $\psi$). Of course, $\bar X $ and
$\bar\psi$ transform with the opposite phases.
\paragraph{Bosonic characters}
For a complex boson $X$ in a twisted sector we get
\bea
\label{comp4}
\chi_{h}^{(X)\,g}(q) & = &
q^{-{1\over 12} + {1\over 2}\nu'(1-\nu')}
\prod_{n=0}^\infty (1 - \ee^{2\pi\ii\nu}q^{n+\nu'})^{-1}
(1 - \ee^{-2\pi\ii\nu}q^{n+(1-\nu')})^{-1}
\nonumber\\
& = & \ii\, \ee^{-\pi\ii\nu} q^{-{\nu'^2\over 2}}
{\eta(\tau)\over\theta_1(\nu+\tau\nu'|\tau)}~.
\eea
In the first line we used the expression of the zero-point energy of a
complex boson in the sector twisted by $h$:
\beq
\label{comp5}
a^{(X)} = {1\over 12} -{1\over 2}\nu'(1-\nu')~,
\eeq
and in the second step we used \eq{theta1}. Notice that when writing infinite
products in terms of $\theta$ functions, $\nu$ and $\nu'$ will always assume
values in the range $[0,1[$.
\par
In the untwisted sector ($h=e$) bosonic zero modes show up. In the closed
string torus compactification, for instance, they are the momentum and winding
modes.  These modes, however, are shared between the chiral and anti-chiral
sectors  and there is no natural split.  Their expression and their effect is
discussed, by means of an explicit  example, in \secn{sec:chiral}. Here we
concentrate on the non-zero-mode  contribution which, for a generic insertion
$g$, reads
\beq
\label{appc6}
\widehat\chi_{e}^{(X)\,g}(q) = 
q^{-{1\over 12}}
\prod_{n=1}^\infty (1 - \ee^{2\pi\ii\nu}q^n)^{-1}
(1 - \ee^{-2\pi\ii\nu}q^n)^{-1}
=  2\sin\pi\nu\,
{\eta(\tau)\over\theta_1(\nu|\tau)}~.
\eeq
\par
The character in the untwisted sector with no insertions coincides with the 
flat space expression. The resulting partition function is of the form
\beq
\label{sera5}
Z(e,e) = Z_{\mathrm{flat}} 
\propto \mathrm{Vol}\, 
{1\over \mathrm{Im}\tau\,|\eta(\tau)|^2}~,  
\eeq
where $1/|\eta(\tau)|^2$ is the contribution of chiral and anti-chiral non-zero
modes while $1/\mathrm{Im}(\tau)$ results from the Gaussian zero-mode
integration (which is replaced by a sum over the appropriate
dual lattice in the compactified case). The flat space partition function is
modular invariant by itself.
\paragraph{Fermionic characters}
Consider now a complex fermion $\psi$ (and its complex conjugate).  Let us
first introduce some more notation,  to distinguish the various fermion
characters. The R sector will be denoted by $\zeta'=0$, while the NS sector
corresponds to  $\zeta'=1/2$; that is, $\zeta'$ corresponds to the shift in the
modings of the oscillators appropriate to the two sectors. Another label $\zeta
= 0 (1/2)$  will denote a Fock space trace with (without) insertion of $(-)^F$.
The insertion of the fermion number $(-)^F$ in the trace is denoted by
$\zeta=0$, while $\zeta=1/2$ denotes the trace without the fermion number
insertion. As an example, $\chi_{{1\over 2},h}^{0,g}$ is the character in the
NS$(-)^F$ sector,  twisted by $h$ and with the insertion of $g$.
\par
The zero-point energies in the  R and NS sectors can be written compactly as
\beq
\label{zeroferm}
a^{(\zeta',\nu')} = 
{1\over 24} -{1\over 2}(\tilde\nu'+\zeta'-{1\over 2})^2~,
\eeq
where $\tilde\nu'$ is defined an integer shift%
\footnote{Remember that the values $\nu'$ and $\nu$ were defined by 
Eqs.~(\ref{comp2}-\ref{comp3}) only up to integer shifts.} of $\nu'$ such that
$\tilde\nu' + \zeta' <1$.  Thus, in the R sector we get   $a^{(\mathrm{R})}=
-a^{(X)} =$ $ -1/12 + \nu'(1-\nu')/2$, while in the NS sector
we have to distinguish
between $\nu'<1/2$ and $\nu'>1/2$. Taking into account \eq{zeroferm}, the
chiral blocks for free fermions are
\beq
\label{fermch}
\chi_{\zeta',h}^{\zeta,g}(q) =  q^{\tilde\nu'^2\over 2}\,
\ee^{-2\pi\ii(\zeta' - {1\over 2})(\nu+\zeta - {1\over 2})}\,
{\theta\sp{1- 2\zeta'}{1 - 2\zeta}(\tau+\tilde\nu'\tau|\tau)\over \eta(\tau)}~.
\eeq
For instance, in the R$(-)^F$ sector, we have 
\bea
\label{RF}
\chi_{0,h}^{0,g}(q)& = & 
q^{{1\over 12} - {\nu'\over 2}(1 -\nu')}\,
\prod_{n=0}^\infty (1 - \ee^{2\pi\ii\n}q^{n+\nu'})
(1 - \ee^{-2\pi\ii\nu}q^{n+ 1 -\nu'})
\nonumber\\
& = &  -\ii\ee^{\pi\ii\nu}\, 
{\theta\sp{1}{1}(\nu+\tau\nu'|\tau)\over \eta(\tau)}~. 
\eea
One may notice that the expression of NS and NS$(-)^F$ characters  twisted by
$\nu'$ coincides (upon use of some theta  function identities)  with the one of
R and R$(-)^F$ characters twisted by $\tilde\nu' + 1/2$. 
\par
Instead of the above set of characters, it is often more useful to use a
different set, labeled by the  $o,v,s,c$ irreducible representations of the
SO$(2)$ algebra.  Taking the relevant combinations of the characters in
\eq{fermch}, one finds
\bea
\label{ovschg}
(\chi_o)_h^g & = &
q^{(\tilde\nu')^2\over 2} \,
{\theta_3(\nu+\tau\tilde\nu'|\tau) + \theta_4(\nu+\tau\tilde\nu'|\tau)
\over 2\eta(\tau)}~,
\nonumber\\
(\chi_v)_h^g & = &
q^{(\tilde\nu')^2\over 2} \,
{\theta_3(\nu+\tau\tilde\nu'|\tau) - \theta_4(\nu+\tau\tilde\nu'|\tau)
\over 2\eta(\tau)}~,
\nonumber\\
(\chi_s)_h^g & = &
q^{(\nu')^2\over 2}\,\ee^{\ii\pi\nu}\,
{\theta_2(\nu+\tau\nu'|\tau) -\ii \theta_1(\nu+\tau\nu'|\tau)
\over 2\eta(\tau)}~,
\nonumber\\
(\chi_c)_h^g & = &
q^{(\nu')^2\over 2} \,\ee^{\ii\pi\nu}\,
{\theta_2(\nu+\tau\nu'|\tau) +\ii \theta_1(\nu+\tau\nu'|\tau)
\over 2\eta(\tau)}~.
\eea
\par
Whenever zero-modes are present, one has to be careful.  In those cases, the
validity of the expressions in \eq{ovschg}  cannot be simply assumed.  Fermion
zero-modes are present in the untwisted R and R$(-)^F$ sectors, and  in the NS
and NS$(-)^F$ sectors twisted by an element of order two (\emph{i.e.}, such that
$\nu'=1/2$). The zero-modes  $e^+=\psi^{i}_0/\sqrt{2}$ and  $e^-=\psi^{\bar
i}_0/\sqrt{2}$, satisfy a Clifford algebra: $\acomm{e^+}{e^-}=1$. This gives a
two-state spectrum containing $\ket\downarrow$, such that
$e^-\ket\downarrow=0$, and $\ket\uparrow=e^+\ket\downarrow$. In this basis, the
rotation generator  for the fermions,   restricted to the zero modes and to
rotations in the 1-2 plane  (having set $\psi^i=\psi^1+\ii\psi^2$) reads simply
$J = J_{12} = {1\over 2} [e^+,e^-] = {1\over 2}\sigma^3$. The action of $g$,
namely a rotation by an angle $2\pi\nu$, is effected on the fermion zero-modes
by $\exp(2\pi\ii\nu J)$, that is by $\cos\pi\nu \mathbf{1} +\ii
\sin\pi\nu\,\sigma^3$. Since the fermion number operator $(-)^F$ is just
$-\sigma^3$, the trace of $g$ over the zero-mode sector with or without the
insertion of the fermion number  operator is given respectively by
$-2\ii\sin\pi\nu$ or by $2\cos\pi\nu$.
\par 
Thus, for instance, in the R$(-)^F$ sector we get,
\beq
\label{comp9bis}
q^{{1\over 12}}\,(-2\ii\sin\pi\nu)\,
\prod_{n=1}^\infty (1 - \ee^{2\pi\ii\nu}q^{n})
(1 - \ee^{-2\pi\ii\nu}q^{n})\nonumber\\
=  -\ii\,{\theta_1(\nu|\tau)\over\eta(\tau)}~,
\eeq
\par
In terms of the $SO(2)$ (generalised) spinorial characters $\chi_{s,c}$, this
translates into the expressions given in \eq{untwferm2} in the text. 
\subsubsection*{Modular transformations}
We outline now the modular properties  of the bosonic and fermionic
characters computed above.
\par
The modular properties of chiral blocks of orbifold theories were described, in
the context of rational  conformal field theories, in~\cite{DVVV}; in these
cases, under the action of the generators $S:\tau\to-1/\tau$ and
$T:\tau\to\tau+1$,  the chiral blocks $\chi_h^g$ transform as follows:
\bea
\label{chirmod}
\chi_h^g & \stackrel{S}{\longrightarrow} & \sigma(h|g)\,\chi_g^{h^{-1}}~;
\\
\chi_h^g & \stackrel{T}{\longrightarrow} & \ee^{-\pi\ii{c\over 12}}\,
\tau_h \,\chi_h^{hg}~.
\eea
The quantities $\sigma(h|g),\tau_h$ are related and encode the 
conformal dimension  $\Delta_h$ of the twist field creating the 
vacuum of the $h$-twisted sector:
\beq
\label{hconf}
\ee^{\pi\ii{c\over 4}}\,\sigma(h|h)^{-1}=\,\tau_e\,(\tau_h)^2 = 
\ee^{4\pi\ii\Delta_h}~.
\eeq  
\par
Since free bosons do not generically build a rational CFT, it is not guaranteed
a priori that the modular properties described above carry over unchanged.
However, by explicitly carrying out the modular transformations relying on the
properties of theta functions, one finds that they do.
\par
The set of bosonic characters given in \eq{comp4} behaves under $S$ as
\beq
\label{modbos}
\chi_{h}^{(X)\,g}(q') = \ii \ee^{2\pi\ii\left(\nu\nu' - {\nu+\nu'\over 
2}\right)} \chi_{g}^{(X)\,h^{-1}}(q)~,
\eeq
where $\tau'=-1/\tau$ and $q'=\exp(2\pi\ii\tau')$, and we used  
\eq{modab} or \eq{modeta}.
This corresponds to
\beq
\label{bosphases}
\sigma(h|g) = \ee^{2\pi\ii(\nu'-{1\over 2})(\nu-{1\over 2})}~.
\eeq
Under the $T$ generator, using theta function properties \eq{modTab} 
and setting $\tau'' = \tau+1$, one has that
\beq
\label{bosT}
\chi_{h}^{(X)\,g}(q'') = \ee^{-\ii{\pi\over 6}}\ee^{\pi\ii\nu'(1-\nu')} 
\chi_{h}^{(X)\,hg}(q)~.
\eeq
Thus, by comparison with \eq{bosT} we get, since $c=2$ for a complex boson,
\beq
\label{tauh}
\tau_h \equiv \ee^{2\pi\ii\Delta_h} = \ee^{\pi\ii\nu'(1-\nu')}~.
\eeq
The quantities  $\sigma(h|g)$ and $\tau_h$ of 
Eqs.~(\ref{bosphases},~\ref{bosT})
correspond to the dimension of the twist field that creates the vacuum of the
$\mathcal{H}_h$ twisted sector being
\beq
\label{boshconf}
\Delta_h= {1\over 2} \nu'(1 - \nu)~, 
\eeq
in agreement  with the expression of the zero-point energy of a complex boson
in the $h$-twisted sector, \eq{comp5}, as  $a^{(X)} = c/24 -\Delta_h$.
\par
We have to consider separately the characters in the untwisted sectors. In the
absence of continuous momentum the untwisted characters are the
$\hat\chi_e^g(q)$ of \eq{appc6},  and the action of $S$, \eq{opmodbos} in the
text, corresponds to
\beq
\label{opbosS}
\widehat\sigma(e|g) = 2\sin\pi\nu~.
\eeq
\par
Let us next consider the fermionic characters \eq{fermch}.  In the twisted
sectors ($h\not= e$), and for $h$ not of order two, under the $S$ modular
transformation we find, using \eq{modsimple},
\beq
\label{modferm}
\chi_{\zeta',h}^{\zeta,g}(q') = 
\ee^{-2\pi\ii(\tilde\nu'+\zeta'-{1\over 2})(\tilde\nu+\zeta-{1\over 2})}\,
\chi_{\zeta,g}^{\zeta',h^{-1}}(q)~,
\eeq
with $\tilde\nu$ being an integer shift of $\nu$ such that $\tilde\nu+1/2<1$,
in analogy with $\tilde\nu'$. This transformations exhibits phases
\beq
\label{din2}
\sigma(\zeta',h|\zeta,g)=
\ee^{-2\pi\ii(\tilde\nu'+\zeta'-{1\over 2})(\tilde\nu+\zeta-{1\over 2})}~.
\eeq
\par
Under the T-modular transformation one finds 
\beq
\label{din1}
\chi_{\zeta',h}^{\zeta,g}(q'')= \ee^{-\pi\ii\over 12} 
\ee^{\pi\ii(\tilde\nu'+\zeta'-{1\over 2})^2}\,
\chi_{\zeta',h}^{\zeta+\zeta'\,\mathrm{mod}\,1,hg}(q)~.
\eeq
These expressions yield conformal weights of the spin-twist fields that agree
with the 0-point energies of \eq{zeroferm}:
\beq
\label{fermhconf}
\tau_{\zeta',h} \equiv \ee^{2\pi\ii\Delta_{\zeta',h}} = \ee^{\pi\ii
(\tilde\nu'+\zeta'-{1\over2})^2}~.
\eeq
The quantities in \eq{din2}  and \eq{fermhconf} correctly satisfy the relation
\eq{hconf}, taking into account the conformal weight $1/8$ of the SO$(2)$ spin
fields that map the NS vacuum to the R vacuum: $\tau_{0,e} = \exp(\pi\ii/8)$.
\par
In the untwisted case, as well as in the in the case in which $h$ is of order
two, the expressions are simpler than those in the generic cases. They turn out
to be
\beq
\label{untwphases}
\sigma(\zeta',e|\zeta,g) = \ee^{-\pi\ii(\zeta'-{1\over 2})(\zeta-{1\over 2})}~,
\eeq
\emph{i.e.}, only the R$(-)^F$ sector acquires a $-\ii$ factor. In the
NS sectors twisted by $h\neq e $ with $h^2=e$ we have the same behaviour, with
\beq
\label{twz2phases}
\sigma(\zeta',h|\zeta,g) = \sigma(\zeta'+{1\over 2}\,\mathrm{mod}\,1,e|\zeta,g)
\hskip 0.4cm \ \ (h^2=e)~.
\eeq
\par
For various purposes, it may be more useful to express the action of the
$S$-modular transformation in terms of the $o,v,s,c$ characters. As such, the
phases are collected in $4\times 4$ matrices  $S(h|g)$ acting on the vector
$(\chi_a)_h^g$, with $a=v,o,s,c$: 
\beq
\label{Shg}
(\chi_a)_h^g \stackrel{S}{\longrightarrow} [S(h|g)]_a^{~b}
(\chi_b)_g^{h^{-1}}~.
\eeq
In terms of the phases of \eq{din2}, we have, for $h$ generic,
\beq
\label{Shgmat}
S(h|g) = {1\over 2} \left(\matrix{
\sigma({1\over 2},h|{1\over 2},g) & \sigma({1\over 2},h|{1\over 2},g) &
\sigma({1\over 2},h|0,g) & \sigma({1\over 2},h|0,g) \cr
\sigma({1\over 2},h|{1\over 2},g) & \sigma({1\over 2},h|{1\over 2},g) &
-\sigma({1\over 2},h|0,g) & -\sigma({1\over 2},h|0,g) \cr 
-\sigma(0,h|{1\over 2},g) & \sigma(0,h|{1\over 2},g) &
\sigma(0,h|0,g) & -\sigma(0,h|0,g) \cr
-\sigma(0,h|{1\over 2},g) & \sigma(0,h|{1\over 2},g) &
-\sigma(0,h|0,g) & \sigma(0,h|0,g)}\right)~.
\eeq
In the untwisted case, the matrix takes a much simpler form. It coincides 
simply with the $S$ matrix for the usual SO$(2)$ chiral blocks, 
given in \eq{ven2} in the text (for $n=1$): 
$S(e|g) = S_{(2)}$.
An analogous expression is obtained when the twist $h$ is of order 2, with 
$(\chi_o)_h^g\leftrightarrow(\chi_s)_e^g$ and 
$(\chi_v)_h^g\leftrightarrow(\chi_c)_e^g$.
\sect{Useful formulae}
\label{app:useful}
%
\subsection{Theta functions}
In terms of the quantities
\bea
\label{qz}
q&=&\exp(2\pi\ii\tau)~;\\
z&=&\exp(2\pi\ii\n)~,
\eea
we define the theta functions with characteristic $\sp ab$
\beq
\label{thetaab}
\theta\sp ab(v|\tau) =
\sum_{n\in\mathbb{Z}} \exp\left\{\ii(n-{a\over 2})^2 \pi\tau
+ 2\pi\ii(v-{b\over 2})(n - {a\over 2})\right\}~.
\eeq
These functions have several periodicity properties rather obvious from their
definition \eq{thetaab}; in particular, 
\beq
\label{shiftb}
\theta\sp{a}{b-2m}(z|\tau) = \ee^{-\pi\ii am} \theta\sp{a}{b}(z|\tau)~,
\hskip 0.4cm m\in\mathbb{Z}~.
\eeq 
If the argument $z$ of the theta-function is of the form $z=\nu +\tau\nu'$ it
can be absorbed in the characteristic vector:
\beq
\label{shifttheta}
\theta\sp ab(\nu + \tau \nu'|\tau) = q^{-{\nu'^2\over 2}}
\ee^{\ii\pi\nu'(b - 2\nu)} \theta\sp{a-2\nu'}{b - 2 \nu}(0|\tau)~.
\eeq
Under the modular transformation $S: \tau \to \tau'=-1/\tau$ we have
\beq
\label{modsimple}
\theta\sp{a}{b}(0|\tau) = (-\ii\tau')^{{1\over 2}} 
\ee^{\ii\pi{ab\over 2}}\,
\theta\sp{b}{-a}(0|\tau')~. 
\eeq
Using \eq{shifttheta} we may also write
\beq
\label{modab}
\theta\sp{a}{b}(\nu+\nu'\tau|\tau) = (-\ii\tau')^{{1\over 2}} 
\ee^{\ii\pi{ab\over 2}}\,
\ee^{\ii\pi\left({(\nu')^2\over\tau'} + \nu^2\tau' - 2\nu'\nu\right)}\,
\theta\sp{b}{-a}(-\nu'+\nu\tau'|\tau')~. 
\eeq
Under the modular transformation $T: \tau\to\tau''$ we have
\beq
\label{modTab}
\theta\sp{a}{b}(\nu+\nu'\tau|\tau) = 
\ee^{\ii\pi{a(a-2)\over 4}}\,
\theta\sp{a}{b-a+1}(\nu-\nu'+\nu'\tau''|\tau'')~. 
\eeq
The theta-functions can be expressed as infinite products by making use of the
``Jacobi triple product identity''
\beq
\label{jactriple}
\prod_{n=0}^\infty (1-q^{n+1})
(1 - wq^{n+{1\over 2}})(1-w^{-1}q^{n+{1\over 2}}) =
\sum_{n\in\mathbb{Z}}q^{n^2\over 2}w^n~.
\eeq
Introducing the usual Jacobi notation
\bea
\label{whichtheta}
&& \theta_1(v|\tau) = \theta\sp 11(v|\tau)~;~~
\theta_2(v|\tau) = \theta\sp 10(v|\tau)~;
\nonumber\\
&&\theta_3(v|\tau) = \theta\sp 00(v|\tau)~;~~
\theta_4(v|\tau) = \theta\sp 01(v|\tau)~,
\eea
one finds
\bea
\theta_1(v|\tau)&=&2\exp(\pi\ii\tau/4)\sin\pi v\prod_{m=1}^\infty(1-q^m)(1-zq^m)
                (1-z^{-1}q^m)~;
                \label{theta1}\\
\theta_2(v|\tau)&=&2\exp(\pi\ii\tau/4)\cos\pi v\prod_{m=1}^\infty(1-q^m)(1+zq^m)
                (1+z^{-1}q^m)~;
                \label{theta2}\\
\theta_3(v|\tau)&=&\prod_{m=1}^\infty(1-q^m)(1+zq^{m-1/2})(1+z^{-1}q^{m-1/2})~;
                \label{theta3}\\
\theta_4(v|\tau)&=&\prod_{m=1}^\infty(1-q^m)(1-zq^{m-1/2})(1-z^{-1}q^{m-1/2})~.
                \label{theta4}
\eea
The theta functions satisfy Jacobi's ``abstruse identity''
\beq
\theta_3(0|\t)^4-\theta_2(0|\t)^4-\theta_4(0|\t)^4=0~.
\eeq
Further,
\beq
\theta_1(0|\t)=0~.
\eeq
The Dedekind eta function is defined by
\beq
\eta(\t)=q^{1/24}\prod_{m=1}^\infty(1-q^m)~.
\eeq
It has the modular transformation properties
\beq
\eta(\t+1)=\exp(\ii\pi/12)\eta(\t)
\eeq
and
\beq
\label{modeta}
\eta(-1/\t)=(-\ii\t)^{1/2}\eta(\t)~.
\eeq
\subsection{Discrete groups}
\label{app:discrete}
In the text we make repeatedly use of the orthogonality properties of the 
character table 
\beq
\label{dg1}
\rho_I^{~\a} \equiv \tr_{\mathcal{D}^I} g^{(\alpha)}
\eeq
of a discrete group $\Gamma$. Above, we have denoted by $\mathcal{D}^I$ the
irreducible representations of $\Gamma$; moreover, we denote by
$\mathcal{C}^\a$ the conjugacy class of the element $g^{(\a)}$ (the traces are
invariant under conjugation). The set of conjugacy classes is in one-to-one 
correspondence with the irreducible representations,  so that the character
table $\rho_I^{~\a}$ is a square matrix.  It enjoys the following orthogonality
properties:
\bea
\label{dg2}
{1\over |\Gamma|}\sum_\a n_\a\, (\rho_I^{~\a})^*\, \rho_J^{~\a} & = &
\delta_{IJ} 
\nonumber\\
{1\over |\Gamma|}\sum_I  (\rho_I^{~\a})^*\, \rho_I^{~\b} & = &
\frac{\delta^{\a\b}}{n_\a}~,
\eea  
where $|\Gamma|$ is the order of $\Gamma$ and $n_\a$ is the number of 
elements of the conjugacy class
$\mathcal{C}^\a$.
%
%

%
\end{document}